# From fields to fuel: analyzing the global economic and emissions potential of agricultural pellets, informed by a case study

Sebastian G. Nosenzo, Rafael Kelman

*PSR Energy Consulting and Analytics, Praia de Botafogo, 370, Botafogo, Rio de Janeiro, RJ, 22250-040, Brazil*

| KEYWORDS | ABSTRACT |
|---|---|
|  | Agricultural residues represent a vast, underutilized resource for renewable energy. This study combines empirical analysis from 179 countries with a case study of a pelletization facility to evaluate the global potential of agricultural pelletization for fossil fuel replacement. The findings estimate a technical availability of 1.44 billion tons of crop residues suitable for pellet production, translating to a 4.5% potential displacement of global fossil fuel energy use, equating to 22 million TJ and equivalent to 917 million tons of coal annually. The economically optimized scenario projects annual savings of \$163 billion and a reduction of 1.35 billion tons of $CO_2$ equivalent in emissions. Utilizing the custom-developed CLASP-P and RECOP models, the study further demonstrates that agricultural pellets can achieve competitive pricing against conventional fossil fuels in many markets. Despite logistical and policy challenges, agricultural pelletization emerges as a scalable, market-driven pathway to support global decarbonization goals while fostering rural economic development. These results reinforce the need for targeted investment, technological advancement, and supportive policy to unlock the full potential of agricultural pellets in the renewable energy mix. |

## 1. Introduction

### 1.1 The Energy Paradigm Shift: Biomass as a Renewable Frontier

An urgent need to decarbonize the global energy system has catalyzed a widespread shift away from fossil fuels, giving rise to two principal avenues of transformation: the electrification of systems through renewable energy sources and the direct substitution of fossil fuels with bio-based alternatives. The former has manifested visibly in the global expansion of wind, solar, and hydroelectric capacity, but the latter remains equally pivotal in the broader pursuit of a low-carbon future [1]. Within this second pathway, biomass emerges not merely as a transitional tool, but as a strategic asset capable of achieving meaningful greenhouse gas mitigation, enhancing energy security, and stimulating localized economic development [2].

Biomass's strategic appeal lies in its dual functionality. First, it addresses energy needs through combustion or conversion, especially in the context of solid biofuels such as pellets for heat and power. Second, it repurposes agricultural waste streams that would otherwise remain underutilized or environmentally burdensome [3]. Historically, the impetus to integrate biomass into the energy portfolio was driven by the 1970s energy crises and concerns over energy self-sufficiency [4]. That urgency has only intensified under the looming shadow of international net-zero pledges soon to reach their deadlines. The bioenergy sector's growth, particularly in solid fuels, has been prompted by volatility in fossil fuel pricing and the intensifying global awareness of climate impacts tied to coal, oil, and natural gas consumption [5]. As such, demand for biomass-based fuels is increasingly driven not only by their capacity to mitigate carbon emissions in the short and medium term, but also for their ability to increase energy security, especially for countries with high fuel import rates [6].

Among these feedstocks, agricultural residues offer unique advantages. Despite their abundance, only a fraction of residues currently see productive use, typically limited to animal bedding or localized soil amendment (See Section 2.3.3). Yet residues that are otherwise discarded—or worse, burned—carry untapped potential for use in solid biofuels, representing a convergence of environmental, economic, and energy benefits [3].

Agriculture itself holds the key to unlocking this potential. Not only does it produce vast quantities of organic byproducts (the global production of primary crops reached 9.9 billion tonnes in 2023 [7]), but productivity gains across major cropping systems can simultaneously support both food and energy needs. As global agriculture seeks to balance rising food demands with the need to curb environmental degradation, a shift toward sustainable bioenergy systems—especially those that valorize existing waste—is not just pragmatic, but imperative [8] [9].

### 1.2 Global Emissions and Energy Potential

Despite decades of international pledges and policy shifts, global greenhouse gas emissions continue to rise, driven largely by fossil fuel combustion in energy, industry, and agriculture. This trend not only exacerbates climate instability but highlights a persistent gap in the global energy transition: the decarbonization of heat and power production, especially in regions reliant on fuel burning. In this context, agricultural residues represent an underutilized yet high-potential energy resource that is readily available, widely distributed, and capable of being converted into solid biofuels that displace more polluting alternatives such as coal, oil, and natural gas.

### 1.3 Global Crop Production and Residue Generation

Modern agriculture produces more than food—it yields an immense byproduct in the form of crop residues. Generated during harvest as stems, stalks, leaves, or empty fruit bunches [3], these represent a latent energy source with global implications. Above-ground crop residues alone account for more biomass than what is harvested as edible grain (see Table S3), and their embodied energy has been estimated at roughly 15% of global primary human energy use [10] [11].

A significant share of these residues remains technically and economically available for energy conversion [5]. Unlike dedicated bioenergy crops, agricultural residues are byproducts that do not compete with food production, offering a rare pathway to expand energy systems without triggering land-use tradeoffs [12]. By redirecting these residues into productive use, particularly through densified fuels like pellets, agricultural systems can shift from being a source of waste to a cornerstone of sustainable energy systems. As global momentum builds toward decarbonization, the scale and structure of crop residue generation make it a foundational resource [13] [14].

### 1.4 Biomass Pelletization

The transformation of loose, heterogeneous biomass into compact, energy-dense pellets has become one of the most promising avenues for renewable energy deployment. Pelletization addresses a fundamental barrier to biomass utilization: its unwieldy form. Raw agricultural and forestry residues often possess high moisture content, irregular size, and low bulk density, making them inefficient to store, transport, or combust in industrial applications. Densification into pellets resolves these logistical challenges by increasing energy density, standardizing particle size, and ensuring year-round availability of feedstock [3] [6] [15].

Pellets streamline the supply chain and outperform traditional biomass in ease of handling, combustion efficiency, and environmental footprint. Their higher bulk density reduces storage space and transportation costs, while their uniformity allows for better combustion in boilers and co-firing with coal [6] [16]. This compatibility with coal-fired power systems makes pellets a powerful transitional fuel, enabling immediate emissions reductions without overhauling existing infrastructure [16].

As energy systems pivot toward decarbonization, the pellet sector has emerged as one of the fastest-growing segments of the bioenergy industry [17] (see Fig. 3). Driven by escalating demand, the market is evolving beyond traditional wood-based inputs to include a broader range of biomass sources, including crop residues, sawdust, and forestry waste [5] [16] [18]. This diversification responds to resource shortages, such as declining availability of sawdust and wood shavings. As such, pelletization now represents not merely a technical process but a strategic intervention: a scalable solution to both energy decarbonization and biomass waste management [3] [5] [15].

### 1.5 Agricultural Residue Pellets

In countries such as India, agricultural residues already form the dominant biomass feedstock for pellet production, underscoring both their accessibility and established use [19]. This trend reflects a broader global shift: as pellet demand rises—propelled by renewable energy targets, market growth, and shortages in traditional wood supply—manufacturers are turning toward agricultural and agro-industrial residues to meet rising consumption. Between 1997 and 2006, pellet market prices surged by approximately 45%, a change that significantly improved the economic viability of using alternative biomass sources, including agricultural by-products [18] [19]. This move is not merely a response to scarcity but a strategic adaptation to ensure supply continuity, especially in regions where wood feedstocks are increasingly diverted to emerging biorefinery markets [20].

Beyond economic drivers, agricultural residues present a critical environmental opportunity. In many developing countries, vast quantities of agro-waste are inefficiently used, openly burned, or simply discarded—practices that waste potential energy resources while contributing to pollution and lost economic value [5]. When densified into pellets, agricultural residues burn significantly cleaner than conventional biomass fuels, achieving carbonaceous particulate matter emissions reductions by 70–90% [21], and offering lower emissions profiles compared to their loose-form counterparts [13]. In stark contrast, the unmanaged burning of crop residues—a practice still prevalent in many agricultural regions—releases massive quantities of particulates and greenhouse gases, worsening both rural and urban air quality [12] [22]. Globally, more than 2 gigatons of crop residues are burned each year, perpetuating a cycle of inefficiency and environmental harm [23] [24].

Using these residues for pellet production addresses both challenges: reducing harmful open-field burning while creating a scalable, decentralized fuel source capable of decarbonizing heat and power generation [25]. Ultimately, agricultural residues are not just a fallback option—they are a necessary evolution in the trajectory of global pellet production. With favorable economics, compatibility with established pelletization infrastructure, and alignment with decarbonization goals, agricultural residues are poised to become a cornerstone of the next generation of biomass energy [3] [18].

*1.6 Scope, Rationale, and Objectives of This Study*

While the viability of biomass pellets has been the subject of multiple studies, existing literature often overlooks a key determinant of real-world applicability: the geographic and crop-specific availability of agricultural residues [3]. Moreover, past analyses frequently generalize feedstock categories without assessing the economic and emissions tradeoffs across different regions or types of residue. This study aims to fill that gap by combining an empirical case study with global quantification models, thereby offering a comprehensive view of agricultural residue pellets as a scalable alternative to fossil fuels.

The rationale behind this investigation lies in the growing need for validated, structured, and evidence-based assessments of renewable energy technologies, especially in a global energy landscape increasingly shaped by sustainability mandates and policy transitions [15]. By modeling the global emissions abatement and economic potential of crop-residue-based pelletization—while grounding the analysis in the technical and logistical realities observed through a case study—this study aims to offer framework for policymakers, investors, and researchers seeking to integrate solid biofuels into the decarbonization agenda.

## 2. Methodology

*2.1 Empirical Case Study of the Pelletization Facility*

This study incorporates a qualitative case analysis of the a pelletization facility in Shanghai, China. The case study approach was selected due to its ability to provide a holistic, context-driven examination of real-world applications, particularly in emerging bioenergy markets where large-scale agricultural pelletization remains underexplored. By analyzing a fully operational pellet production system, this study seeks to ground theoretical insights in operational realities and elucidate agricultural pelletization's broader implications for global biomass utilization.

Data were collected through direct field observations, structured interviews with factory personnel, and a review of internal operational records. The site visit enabled a detailed examination of the facility's infrastructure, production processes, and feedstock management, offering insights into the logistical and technical factors influencing pelletization. Rather than serving as a basis for quantitative modeling, the case study informs the broader discussion on practical challenges, operational structures, and the contextual factors shaping pellet production at the local level.

Additionally, interviews with key stakeholders, including factory management, technical staff, and affiliated local farmers, provided qualitative perspectives on market conditions, policy impacts, and supply chain dynamics. This combination of empirical observation and direct stakeholder engagement ensures a robust analytical foundation, facilitating an evidence-based discussion on the scalability of agricultural pelletization as a commercially viable and environmentally responsible energy solution. Ethical guidelines were rigorously observed throughout the research process, with explicit consent from factory management secured prior to data collection and analysis [26].

*2.2 Wood Pellet Trends as a Model for Agricultural Pellets*

The rapid growth of the wood pellet industry over the past decade provides a compelling precedent for evaluating the potential of agricultural residues in pellet production. As the most widely produced and commercialized form of solid biofuel, wood pellets that are primarily derived from sawdust have become a cornerstone of the global bioenergy market [6]. Their success highlights how standardization, supply chain development, and growing demand can transform a once-niche fuel into a scalable, international commodity.

To assess these dynamics, global wood pellet production data from 2012 to 2022 was analyzed, drawing on official statistics from UN Data/FAOSTAT [27]. The analysis examined year-on-year production changes, regional trends, and overall global growth patterns. These insights offer a valuable lens through which to explore the scalability and market potential of agricultural residue pellets, positioning wood pelletization trends as a reference point for emerging biomass markets.

*2.3 Study Area and Quantification of Residue Availability*

This study quantifies the global availability of agricultural residues suitable for pelletization, establishing a data-driven foundation for evaluating both emissions reduction and economic potential. Using crop production data for 178 countries, sourced from UN Data/FAOSTAT, this study modeled the generation of agricultural residues by applying crop-specific residue-to-product ratios, adjusted for collection efficiency and competing uses [19]. This approach allows for a comprehensive, country-level analysis of theoretical residue availability worldwide.

The selection of wheat, maize, rice, and sugarcane as focal crops reflects both their dominance in global agriculture and their prominence as sources of lignocellulosic residues suitable for bioenergy. Together, residues from these four crops—wheat straw, rice straw, maize stover, and sugarcane bagasse—account for the vast majority of agricultural byproducts available for biomass generation [12] [28-30]. Estimates place their combined residue production at over 1.4 billion tons annually, underscoring their strategic relevance to any global biomass energy initiative. These residues are not only abundant but also structurally suited for pelletization, with high cellulose and hemicellulose content critical for solid fuel applications [30].

Beyond their sheer volume, the geographic distribution of these crops enhances their appeal for global energy modeling. Wheat and rice residues are predominantly produced in Asia, while maize and sugarcane residues are concentrated in the Americas, offering a balanced, cross-regional biomass supply [12]. Their selection also aligns with existing literature identifying these four residue types as the most viable for large-scale biomass generation due to their physical properties and processing potential [31] [32].

Despite their availability, most crop residues remain underutilized for energy production, often left in the field or repurposed for soil maintenance—a practice vital for soil health but leaving significant surplus unexploited [33]. By quantifying this unclaimed fraction, this study aims to assess the true global potential of agricultural residues as a sustainable, scalable energy resource.

### 2.3.1 Global Crop Production

To establish a consistent baseline for residue availability, country-level production data for maize, rice, sugarcane, and wheat in 2021, expressed in metric tons, was compiled. Official statistics were sourced from UN Data/FAOSTAT [34-37], providing a comprehensive and standardized global dataset for the 178 countries analyzed.

### 2.3.2 Available Crop Residue

To estimate the share of crop residues practically available for pelletization, this study applied a multi-step filtering approach that accounts for biological constraints and processing realities.

The residue-to-crop ratio defines the amount of agricultural residue generated per unit of crop production. This ratio varies by crop and forms the first filter in quantifying theoretical residue availability.

The sustainable removal rate (SRR) reflects the proportion of crop residues that can be harvested without compromising soil health and productivity, as unregulated removal of crop residues can degrade soil health and cause enduring environmental harm. For most crops, a balance requires leaving a significant portion of residues on the field to maintain soil organic matter levels, recycle nutrients, stabilize soil structure and tilth, lower soil bulk density, enhance water retention and movement, sustain microbial activity, and prevent erosion [4] [38].

Based on this principle, Scarlat et al. identified average sustainable removal rates for key crops such as maize, wheat, and rice, applying agronomic and environmental considerations to define these thresholds [11]. Sugarcane residues require a differentiated approach: while bagasse—the fibrous by-product of crushing—can be fully collected during processing, only a controlled fraction of the leaves and tops removed at harvest should be collected to protect soil quality [39] [40]. Accordingly, this study applies crop-specific SRRs, recognizing both agronomic sustainability and operational practices in defining the residue share available for pelletization.

In addition, the moisture content of residues directly affects their usable weight for energy applications. The dry matter ratio adjusts the total residue weight to reflect its moisture content, following standard practice in biofuel analyses where most moisture is measured on a dry basis [41].

Tables 1 and 2 present the residue-to-production ratios and the sustainable removal rates applied for each crop in this study—key parameters that shape the theoretical availability of residues for energy use.

Table S2 further details the dry matter ratio applied per country, reflecting climatic and agronomic variations affecting residue moisture content [42]. World average dry matter ratio values were applied for countries with no dry matter ratio data

To estimate the crop residue available for use, the following model was used (Eq. 1-3).

$$CR_a = \left(CR_{total} - CR_{lf}\right) \cdot DMR \qquad (1)$$

where $CR_a$ are the crop residues that is available from removal from agricultural fields, $CR_{total}$ are the total crop residues resulting from agricultural production, $CR_{lf}$ are the crop residues left on agricultural fields, and $DMR$ is the dry matter ratio of the crop residues. For countries with no data, a world average was used.

$$CR_{total} = P \cdot RTP \qquad (2)$$

where $CR_{total}$ are the total crop residues resulting from agricultural production, $P$ is the crop production, and $RTP$ is the residue-to-production ratio.

$$CR_{lf} = CR_{total} \cdot SRR \qquad (3)$$

where $CR_{lf}$ are the crop residues left on the fields, $CR_{total}$ are the total crop residues resulting from agricultural production, and $SRR$ is the sustainable removal rate of crop residues from agricultural fields.

**Table 1.** Aboveground crop residue as a percentage of crop production (residue-to-production)

| Crop | Value | Unit | Reference |
|------|-------|------|-----------|
| Maize | 100 | % | [43] |
| Rice | 140 | % | [43] |
| Sugarcane | 100 | % | [43] |
| Wheat | 130 | % | [43] |

**Table 2.** Percentage of aboveground crop residues that can be sustainably removed from agricultural fields (sustainable removal rate)

| Crop | Value | Unit | Reference |
|------|-------|------|-----------|
| Maize | 50 | % | [11] |
| Rice | 60 | % | [11] |
| Sugarcane Tops | 75 | % | [40] |
| Sugarcane Bagasse | 100 | % | [40] |
| Sugarcane (Overall) | 87.5 | % | [39] [40] |
| Wheat | 40 | % | [11] |

### 2.3.3 Current Crop Residue Utilization

Of the crop residues technically available for pelletization, a portion is already diverted to other uses, reducing the share technically "available" for energy production. Both Smerald et al. and Scarlat et al. identify energy production and livestock feed or bedding as the predominant current uses of crop residues, with Scarlat et al. specifying cattle, horses, sheep, and swine as the main livestock categories utilizing residues for feed or bedding [10] [11]. <u>Table S4</u> details livestock numbers by country [44-47], and <u>Table 3</u> outlines the estimated daily consumption of crop residues by each livestock type, providing a basis for calculating national-level feed demand.

The current bioenergy use of agricultural residues was estimated using UN Data/FAOSTAT statistics on bioenergy production by country [48]. Reported figures were disaggregated into sugarcane bagasse and "other vegetal residues." To attribute a portion of these "other residues" to wheat, rice, and maize, the analysis applied a two-step adjustment: first, considering that cereals account for 31.3% of total primary crop production [7], and second, recognizing that wheat, rice, and maize represent approximately 91% of global cereal output [49]. This method enabled a conservative estimation of the current utilization of these key residues in bioenergy production.

**Table 3.** Estimated per animal crop residue used for animal feed/bedding

| Animal | Val. | Unit | Reference |
|--------|------|------|-----------|
| Cattle | 0.375 | kg/day | [10] |
| Horses | 1.500 | kg/day | [10] |
| Sheep | 0.100 | kg/day | [10] |
| Swine | 0.063 | kg/day | [10] |

### 2.3.4 Final Crop Residue for Pelletization

To estimate the final amount of crop residues available for pelletization, the following model was used (Eq. 4).

$$CR_f = CR_a - \sum_{cu=1}^{n} CR_{cu} \qquad (4)$$

where $CR_f$ are the final crop residues available for pelletization, $CR_a$ are the crop residues that is available from removal from fields, $n$ is the number of current uses for crop residues, and $CR_{cu}$ are the current uses of crop residues, including for bioenergy and for animal feed/bedding.

### 2.4 Available Potential Agricultural Pellet Energy

The potential pellet energy was derived from the final available crop residues identified using the average lower heating values (LHV) of the four selected crops, weighted according to each crop's share in a country's total available residue pool. Additionally, an efficiency adjustment was applied to account for material losses during the pelletization process, following operational data outlined in [26].

### 2.5 Economic & Emissions Modeling

Building on the country-level estimates of available crop residues, this study performed a global analysis of the economic viability and emissions reduction potential of agricultural pelletization across 178 countries. For each country, capital expenditure (CAPEX) and operational expenditure (OPEX) were estimated to establish baseline economic conditions. Two specialized models were then applied using country-specific data: the Country-Level Analysis of Selling Price for Pellets (CLASP-P) and the Replacement Emissions and Cost-Optimizing Planner (RECOP). These models, detailed in the following sections, allowed for the derivation of country-specific results, providing a comprehensive assessment of both economic performance and potential emissions savings associated with agricultural residue pelletization.

### 2.5.1 Capital & Operating Expenditure

Capital expenditure (CAPEX) and operating expenditure (OPEX) estimates for agricultural pelletization were based on a 2023 study by Sarker et al., which modeled an 40,080 t/y agricultural pellet facility in Saskatchewan, Canada over 19 years [50]. To account for regional cost differences, Price Level Indexes (PLIs) were applied to adjust for each country across key variables: labor, raw materials, construction, and electricity [51-54]. In cases where country-specific PLIs were unavailable, average values by continent were used as adjustment coefficients. Tables 4 and 5 present the reference CAPEX and OPEX values applied in this study, serving as the basis for country-level economic modeling.

To calculate country specific capital expenditure (CAPEX) values, the following model (Eq. 5) was used in combination with reference values from Table 4.

$$EPC_c = EPC_{ref} \cdot I_{ref, c, con} \qquad (5)$$

where $EPC_c$ is the equipment purchase cost for country $c$, $EPC_{ref}$ is the reference equipment purchase cost, and $I_{ref, c, con}$ is the index ratio between the reference value and value for country $c$ for construction.

To calculate country specific operating expenditure (OPEX) values, the following model (Eq. 6) was used in combination with reference values from Table 5.

$$OPEX_c = RM_{ref} \cdot I_{ref, c, rm} + L_{ref} \cdot I_{ref, c, labor} + U_{ref}$$
$$\cdot I_{ref, c, elec} + M_{ref} \cdot I_{ref, c, con} + IT_{ref} + AE_{ref} \qquad (6)$$

where $OPEX_c$ is the operating expenditure for country $c$, $RM_{ref}$ is the reference raw material cost, $I_{ref, c, rm}$ is the index ratio between the reference value and value for country $c$ for raw materials, $L_{ref}$ is the reference labor cost, $I_{ref, c, labor}$ is the index ratio between the reference value and value for country $c$ for labor, $U_{ref}$ is the reference utility cost, $I_{ref, c, elec}$ is the index ratio between the reference value and value for country $c$ for electricity, $M_{ref}$ is the reference maintenance cost, $I_{ref, c, con}$ is the index ratio between the reference value and value for country $c$ for construction, $IT_{ref}$ is the reference insurance and tax cost, and $AE_{ref}$ is the reference additional expenses cost.

**Table 4.** Reference values adapted from [50] for CAPEX estimation

| Description | Value | Unit |
|---|---|---|
| Equipment purchase cost (1) | 1,249,570 | US$ |
| Total plant direct cost (2) | 325 | % of (1) |
| Total plant indirect cost (3) | 22 | % of (2) |
| Miscellaneous (4) | 10 | % of (2)+(3) |
| Total fixed capital cost (5) | (2)+(3)+(4) | - |
| Working capital (6) | 5 | % of (5) |
| Start-up cost (7) | 15 | % of (5) |
| CAPEX | (5)+(6)+(7) | - |

**Table 5.** Reference values adapted from [50] for OPEX estimation

| Description | Value | Unit |
|---|---|---|
| Raw material | 482,600 | US$/y |
| Labor | 812,800 | US$/y |
| Labor (Overhead) | 558,800 | US$/y |
| Labor (Supervision) | 152,400 | US$/y |
| Utilities | 203,200 | US$/y |
| Maintenance | 152,400 | US$/y |
| Insurance & taxes | 101,600 | US$/y |
| Additional Expenses | 76,200 | US$/y |

### 2.5.2 CLASP-P

The Country-Level Analysis of Selling Price for Pellets (CLASP-P) is a minimum selling price optimization model specifically designed in this study for agricultural pellet facilities. The model uses inputs of country-specific CAPEX and OPEX, along with a chosen study period of 20 years and an annual output of 40,080 tons per year—a reference value derived from Sarker et al. [50]. The target net revenue for this study was set at $0, with the objective of identifying the minimum selling price required for the facility's net present value (NPV) to reach zero over the designated study period. CLASP-P iteratively adjusts the selling price to optimize for this breakeven NPV, providing country-specific price thresholds under varying economic conditions.

CLASP-P utilizes the following models (Eq. 7-14).

$$MSP = P \mid NPV = 0 \tag{7}$$

where $MSP$ is the minimum selling price, $P$ is the price of pellets, and $NPV$ is the net present value.

$$NPV = \sum_{t=1}^{n} \frac{CF_t}{(1+r)^t} + \frac{S}{(1+r)^n} - CAPEX \tag{8}$$

where $NPV$ is the net present value, $t$ is the year, $n$ is the total number of years, $CF_t$ is the cash flow in year $t$, $r$ is the estimated discount rate for each country, $S$ is the amount salvaged, and $CAPEX$ is the capital expenditure. For countries with no data for the relevant discount rate, a continent-wide average was used.

$$CF_t = R - OPEX - T \tag{9}$$

where $CF_t$ is the cash flow in year $t$, $R$ is the revenue in year $t$, OPEX is the operating expenditure in year $t$, and $T$ is the tax amount in year $t$.

$$R = P \cdot Q \tag{10}$$

where $R$ is the revenue in year $t$, $P$ is the price of pellets, and $Q$ is the output quantity of pellets in year $t$, in this study 40,080 t/y [50].

$$T = TR \cdot (R - OPEX - D) \tag{11}$$

where $T$ is the tax amount in year $t$, $TR$ is the estimated tax rate in each country, $R$ is the revenue in year $t$, OPEX is the operating expenditure in year $t$, and $D$ is the depreciation in year $t$. For countries with no data for the relevant corporate taxation rate, a continent-wide average was used.

$$D = \frac{TFC - S}{n} \tag{12}$$

where $D$ is the depreciation in year $t$, $TFC$ is the total fixed capital cost for the plant, $S$ is the salvage amount, and $n$ is the total number of years.

$$S = SR \cdot TFC \tag{13}$$

where $S$ is the salvage amount, $SR$ is the salvage rate, in this study 10% [50], and $TFC$ is the total fixed capital cost for the plant.

$$TFC = R_{TFC/CAPEX} \cdot CAPEX \tag{14}$$

where $TFC$ is the total fixed capital cost for the plant, $R_{TFC/CAPEX}$ is the ratio between the total fixed capital cost and capital expenditure, and $CAPEX$ is the capital expenditure.

### 2.5.3 RECOP

The Replacement Emissions and Cost-Optimizing Planner (RECOP) is a multi-scenario optimization tool specifically designed in this study that operates with inputs of a country and a price per ton for pellets—set in this study as the minimum selling price derived from the CLASP-P model for that same country under the aforementioned conditions. RECOP then pulls country-specific data on fossil fuel prices (coal, oil, and natural gas), national energy consumption of these fossil fuels, and the amount of available energy from agricultural pellets, as calculated in this study. National fossil fuel price data is estimated by identifying the key fossil fuel price indexes relevant to each continent [55-64], calculating a continental average based on these indexes, and applying this average to individual countries. The index prices are derived from [65-70]. Finally, the model requires a scenario selection between A (Economically Optimized), B (Emissions Optimized), and C (Economically Optimized with User-Defined Carbon Tax). RECOP then ranks coal, oil, and natural gas based on the selected scenario—prioritizing the fuel with the highest potential economic savings (Scenario A), greatest emissions reduction (Scenario B), or highest savings considering a carbon tax (Scenario C). The model then allocates pellet energy starting with the highest-ranked fuel. If pellet supply exceeds the replacement potential of the top-ranked fuel, the surplus is allocated sequentially to the next-ranked fuels until all available pellet energy is used or all national fossil fuel demand is offset.

For this study, Scenario A (Economically Optimized) results were used for each country, reflecting the primary objective of assessing the economic viability of agricultural pellet fuel as a replacement for fossil fuels.

### 2.5.4 Quantification of Economic & Emissions Savings

The outputs generated by the RECOP model serve as the basis for quantifying both the economic benefits and emissions reduction potential associated with agricultural pellet utilization. For each country, RECOP's optimization results—reflecting the allocation of pellet energy to replace coal, oil, and natural gas—determine the volume of fossil fuel displacement and corresponding cost implications.

To accurately calculate both economic savings and avoided emissions, the lower heating value (LHV) and emissions factor (EF) for each fossil fuel type, as well as for agricultural pellets was considered. These parameters allowed for energy content normalization and consistent comparison of emission outputs across fuels. <u>Tables 6</u> and <u>7</u> present the LHV and EF values applied in this analysis.

In addition to direct economic savings and emissions reduction, this study also evaluates the levelized cost of energy (LCOE) for each fossil fuel and for agricultural pellets as an additional measure of economic potential. By assessing the cost per unit of useful energy output over the operational life of each fuel type, the LCOE analysis provides a comparative measure of long-term economic competitiveness across energy sources. Importantly, the analysis reflects the operational advantage of pellets in existing energy systems. Given their compatibility with direct combustion and co-firing in coal pulverization systems, pellets offer a cost-effective pathway for reducing $CO_2$ emissions from coal-fired power plants [16]. This compatibility underscores their practical potential for integration into existing industrial energy infrastructures, enhancing both economic feasibility and environmental impact.

To calculate the levelized cost of energy (LCOE) for each fuel in each country, the following model was used (Eq. 15).

$$LCOE_f = \frac{P_f}{LHV_f} \qquad (15)$$

where $LCOE_f$ is the levelized cost of energy using fuel $f$, $P_f$ is the price of fuel $f$, and $LHV_f$ is the lower heating value for fuel $f$.

To calculate the levelized cost of energy (LCOE) of pellets for each country, the following model was used (Eq. 16).

$$LCOE_p = \frac{P_p}{\overline{x}_w(LHV_{crop})} \qquad (16)$$

where $LCOE_p$ is the levelized cost of energy using crop residue pellets, $P_p$ is the price of crop residue pellets, and $\overline{x}_w(LHV_{crop})$ is the weighted mean of the lower heating values for maize, rice, sugarcane, and wheat residues.

Values for levelized cost of energy are then converted from \$/TJ to \$/MWh by a standard multiplication factor of 277.78 MWh/TJ.

To calculate economic and emissions savings for each country, the following models were used (Eq. 17-18).

$$S_{ec} = \sum_{f=1}^{3} NRG_f \cdot \left(P_f - P_p\right) \qquad (17)$$

where $S_{ec}$ is the economic savings, $NRG_f$ is the energy energy replaced by pellets for fuel $f$, $P_f$ is the price of fuel $f$, and $P_p$ is the price of pellets.

$$S_{em} = \sum_{f=1}^{3} NRG_f \cdot \left(EF_f - EF_p\right) \qquad (18)$$

where $S_{em}$ is the emissions savings, $NRG_f$ is the energy energy replaced by pellets for fuel $f$, $EF_f$ is the emission factor for fuel $f$, and $EF_p$ is the emission factor for pellets.

**Table 6.** Lower Heating Value (LHV) of crop residues/fuels

| Crop/Fuel | Value | Unit | Reference |
|-----------|-------|------|-----------|
| Maize | 17.3 | MJ/kg | [71] |
| Rice | 14.6 | MJ/kg | [71] |
| Sugarcane | 17.3 | MJ/kg | [71] |
| Wheat | 17.2 | MJ/kg | [41] |
| Coal | 23.9 | MJ/kg | [72] |
| Oil | 42.0 | MJ/kg | [72] |
| Natural Gas | 42.0 | MJ/kg | [72] |

**Table 7.** Emissions Factor (EF) of fuels

| Fuel | Value | Unit | Reference |
|------|-------|------|-----------|
| Pellets | 151 | kgCO$_2$e/t | [73-75] |
| Coal | 2592 | kgCO$_2$e/t | [76] |
| Crude Oil | 2977 | kgCO$_2$e/t | [76] |
| Natural Gas | 2114 | kgCO$_2$e/t | [76] |

# 3. Case Study: Agricultural Pelletization Model

## 3.1 Contextualizing the Pelletization Facility

As China accelerates its transition toward renewable energy and rural economic revitalization, the country has articulated national targets for 2030: peak carbon emissions, a 60–65% reduction in carbon intensity relative to 2005 levels, and a 20% share of non-fossil energy in its primary energy mix—all framed within a sustainable development strategy that aligns both national imperatives and international commitments [77]. These objectives have catalyzed a wave of decentralized energy initiatives, especially those that link renewable energy production with rural economic development.

The pelletization facility in question emerged within this policy environment in 2021, as both a commercial enterprise and as a rural development initiative stemming from local government efforts to promote sustainable waste management and stimulate rural economic development in the region. The facility's primary mission is to transform agricultural residues—primarily straw and pruning waste—into biomass pellets for energy use. Beyond its production role, the project aims to establish a closed-loop agricultural waste management system by integrating collection, processing, and local market distribution. Its operational model reflects both environmental objectives and economic goals of increasing rural income through waste valorization. In this dual role, the facility serves as a practical case of how agricultural residue projects like a pellet factory can contribute to broader economic and decarbonization targets.

## 3.2 Technical Workflow of Pellet Production

The workflow at said pelletization facility begins with the collection of raw biomass, which is transported from local farms to the facility's intake area. Upon arrival, the biomass undergoes a preliminary inspection and is sorted to remove unsuitable materials.

The operational steps of pelletization into uniformly sized solid pieces of high bulk density which can be conveniently used as a fuel [5], include mechanical crushing, which reduces the raw biomass to a uniform particle size suitable for pelletization; drying, to achieve the optimal moisture content required for stable pellet formation; and compression under high pressure using specialized pellet mills. This process not only shapes the material into standardized pellets but also increases its density and energy content. The pellets are then cooled to stabilize their structure and prevent degradation during storage. Final processing includes packaging or bulk storage, depending on market demand and distribution plans.

Throughout the operation, the facility emphasizes process efficiency and minimal environmental impact, ensuring that the production line remains both technically viable and aligned with local sustainability objectives. While the system is designed to handle agricultural residues of varying composition, operational adjustments—such as drying time or pressure settings—are made to maintain consistent product quality. Underline{Figures 1} and 2 illustrate the pelletization process, relevant machinery, and final products at the pelletization facility.

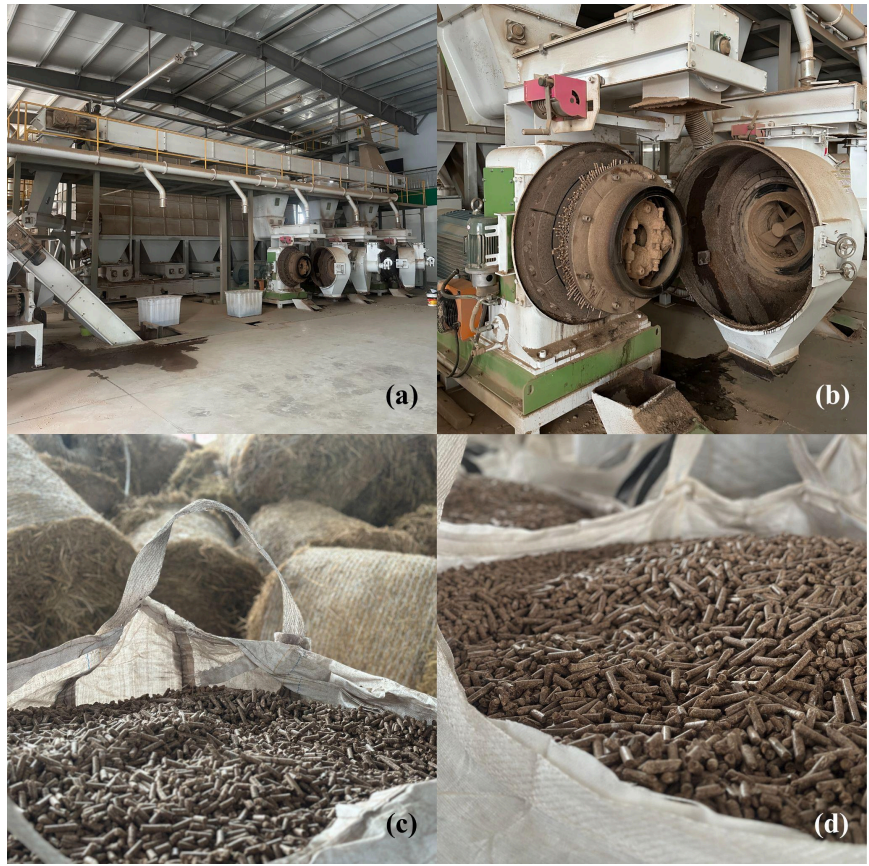

**Fig 1.** Images from the studied pelletization facility, of the pelletization mechanism (a), pellet extrusion machinery (b), agricultural residue/finalized pellet products (c), and collection/organization of agricultural pellets (d).

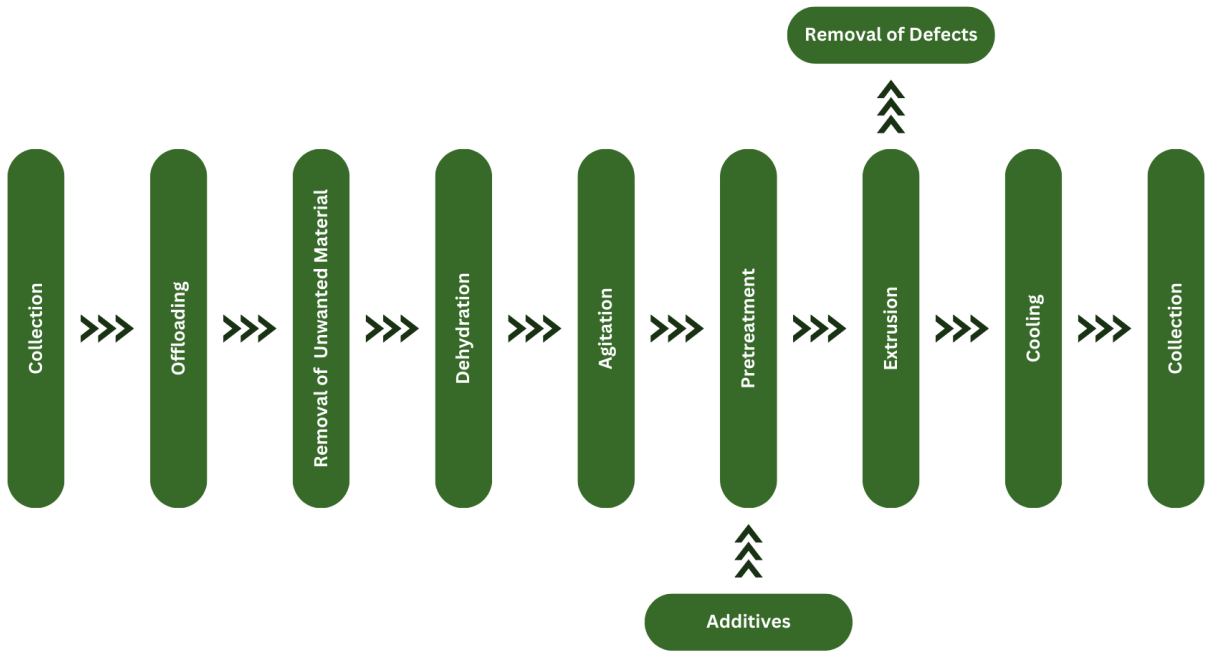

**Fig 2.** Pelletization process, adapted from [26].

## 3.3 Projected Impacts: Emissions, Economic, and Social Co-Benefits

Building on its operational model, the pelletization facility demonstrates tangible economic, environmental, and social outcomes that reflect the broader potential of agricultural residue pelletization. The facility operates with an annual production of 20,000 tons of biomass briquette fuel and 10,000 tons of biomass pellet fuel, generating an estimated 13.3 million yuan in annual sales. After accounting for production and operating costs of approximately 10.6 million yuan, the facility achieves a gross annual profit of 2.7 million yuan.

A defining feature of the facility's model is its "1+2+N+1" approach, a strategy designed to streamline collection, processing, and distribution while embedding community participation at each stage. Under this model, one leading enterprise partners with two central service platforms (environmental sanitation operations and agricultural recycling cooperatives), supplemented by numerous farmer participants ("N"), with oversight and coordination from a township-level management office ("1"). This structure ensures comprehensive coverage of residue collection, leverages existing sanitation infrastructure for transportation, and promotes shared economic benefits across the supply chain.

The facility processes approximately 20,000 tons of rice and wheat straw annually, incentivized by government subsidies of about 100 yuan per ton. Additionally, it handles 15,000 tons of fruit and vegetable vines, collected through an environmental sanitation bidding process at 80 yuan per ton. These feedstocks are integrated into a system which consolidates various agricultural residues into a single processing stream.

Economically, the facility's product fuel production achieves a calorific value of 3,000–3,400 kcal, with a bulk density between 0.4–0.6 t/m³. The production cost averages 220 yuan per ton, while the market price stands at 320 yuan per ton, yielding a net benefit of approximately 100 yuan per ton. The operation also plans to expand into cattle and sheep feed production, targeting a profit margin of 320 yuan per ton based on a projected production cost of 980 yuan per ton and a market price of 1,300 yuan per ton.

From an emissions perspective, the facility plays a direct role in reducing agricultural waste disposal through structured collection and processing. By handling approximately 35,000 tons of crop straw and greening branches annually, the operation helps alleviate environmental pressures, ease resource constraints, and maintain agricultural ecological balance in the region.

Beyond waste management, the facility produces biomass RDF fuel, offering a renewable energy alternative that supports circular resource use. Its annual output of pellets and briquettes is estimated to offset the equivalent of more than 20,000 tons of standard coal, contributing meaningfully to national carbon reduction goals while promoting sustainable energy recycling.

Socially, the pelletization facility contributes to both rural livelihood improvement and community development. By enhancing environmental services and agricultural waste management, the project improves the production conditions, living standards, and overall ecological environment for farmers in the region.

In addition to environmental benefits, the model fosters community engagement by embedding farmers within its collection and supply network, ensuring that economic benefits extend beyond the enterprise to rural households.

The facility itself strengthens the village collective economy, generating approximately 1 million yuan in annual income for the collective. These outcomes support the region's broader ambition of becoming a benchmark town for rural revitalization and common prosperity.

This case highlights how an integrated agricultural pelletization facility can simultaneously deliver emissions reduction, generate economic returns, and support social development objectives. These observed dynamics offer a grounded reference point for evaluating the scalability of agricultural pelletization in diverse global contexts.

## 4. Results

### 4.1 Wood Pellet Trends

In 2022, global wood pellet production reached 46.4 million tons, reflecting the continued expansion of the biomass pellet market. The leading producers at the country level were the United States of America (9.5 million tons), Canada (3.8 million tons), Germany (3.6 million tons), Vietnam (3.5 million tons), and the Russian Federation (2.8 million tons). These countries collectively accounted for a significant share (50%) of global output, underscoring their central role in the international pellet trade.

Between 2012 and 2022, the countries with the highest year-on-year growth in wood pellet production were Australia (520.5%), Montenegro (301.8%), Thailand (256.9%), Argentina (152.9%), and Vietnam (79.2%). This rapid growth in emerging markets suggests a widening geographic diversification of production, alongside sustained increases in established markets.

Figure 3 presents the global production of wood pellets and the year-on-year growth for 60 countries with available data from UN Data/FAOSTAT between 2012 and 2022, highlighting both dominant producers and emerging markets.

Overall, wood pellet production trends indicate a robust and expanding global industry. The growth reflects rising demand for renewable solid biofuels, increasing adoption of biomass in energy mixes, and the strategic role of pellets as a transitional fuel in both industrial and residential sectors.

### 4.2 Residue Availability by Crop and Country

In 2021, the combined global production of maize, rice, sugarcane, and wheat totaled 4.6 billion tons, underscoring their dominant role in global agriculture. Individually, production volumes reached 1.2 billion tons for maize, 787.3 million tons for rice, 1.9 billion tons for sugarcane, and 770.9 million tons for wheat. For maize, the leading producers were the United States (383.9 million tons), China (272.8 million tons), Brazil (88.5 million tons), Argentina (60.5 million tons), and Ukraine (42.1 million tons). In the case of rice, production was dominated by China (214.4 million tons), India (195.4 million tons), Bangladesh (56.9 million tons), Indonesia (54.4 million tons), and Vietnam (43.9 million tons).

Sugarcane recorded the highest global production among the four crops, driven primarily by Brazil (715.7 million tons), India (405.4 million tons), China (107.3 million tons), Pakistan (88.7 million tons), and Thailand (66.3 million tons). Notably, despite its massive output, sugarcane is cultivated in a comparatively smaller number of countries, making its production highly concentrated geographically. For wheat, the top producers were China (137.0 million tons), India (110.0 million tons), Russian Federation (76.1 million tons), United States (44.8 million tons), and France (36.6 million tons).

Table S1 outlines the total crop production for maize, rice, sugarcane, and wheat by country. These production figures highlight both the global scale and regional concentration of major crop outputs, with Asia and the Americas emerging as pivotal regions for residue generation. The high production volumes—particularly of maize and sugarcane—suggest significant potential for residue availability, reinforcing the relevance of these crops for global biomass supply.

The total crop production figures, however, do not directly translate to biomass available for pelletization. Actual residue availability depends on crop-specific residue-to-product ratios. Table S3 and Figure 4 outline the total crop residues produced for maize, rice, sugarcane, and wheat by country, based on these agronomic parameters.

Further refining this potential, sustainable removal rates and dry matter ratios are considered, yielding the final volume of residues theoretically available for energy applications. From this pool, a significant share is already diverted to competing uses—primarily livestock feed and bedding, as well as bioenergy generation within existing industrial systems.

In 2021, the global use of crop residues for animal feed and bedding amounted to 311.4 million tons, reflecting its role as a key agricultural input. The leading utilizers were Brazil (35.6 million tons), India (29.5 million tons), China (27.5 million tons), United States (20.6 million tons), and Ethiopia (11.6 million tons).

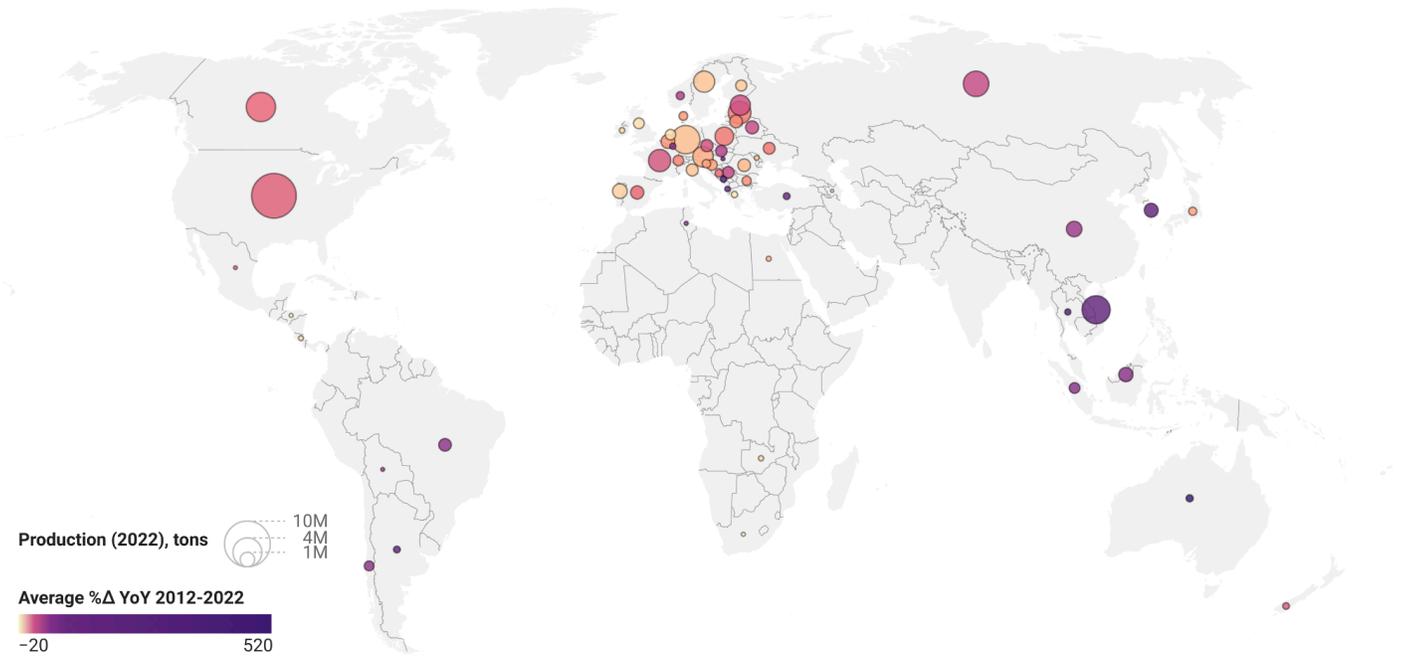

**Fig 3.** Wood pellet production (2022), in tons. Color scale represents average year-on-year percentage change in pellet production, 2012-2022, interpolated in deciles, by country.

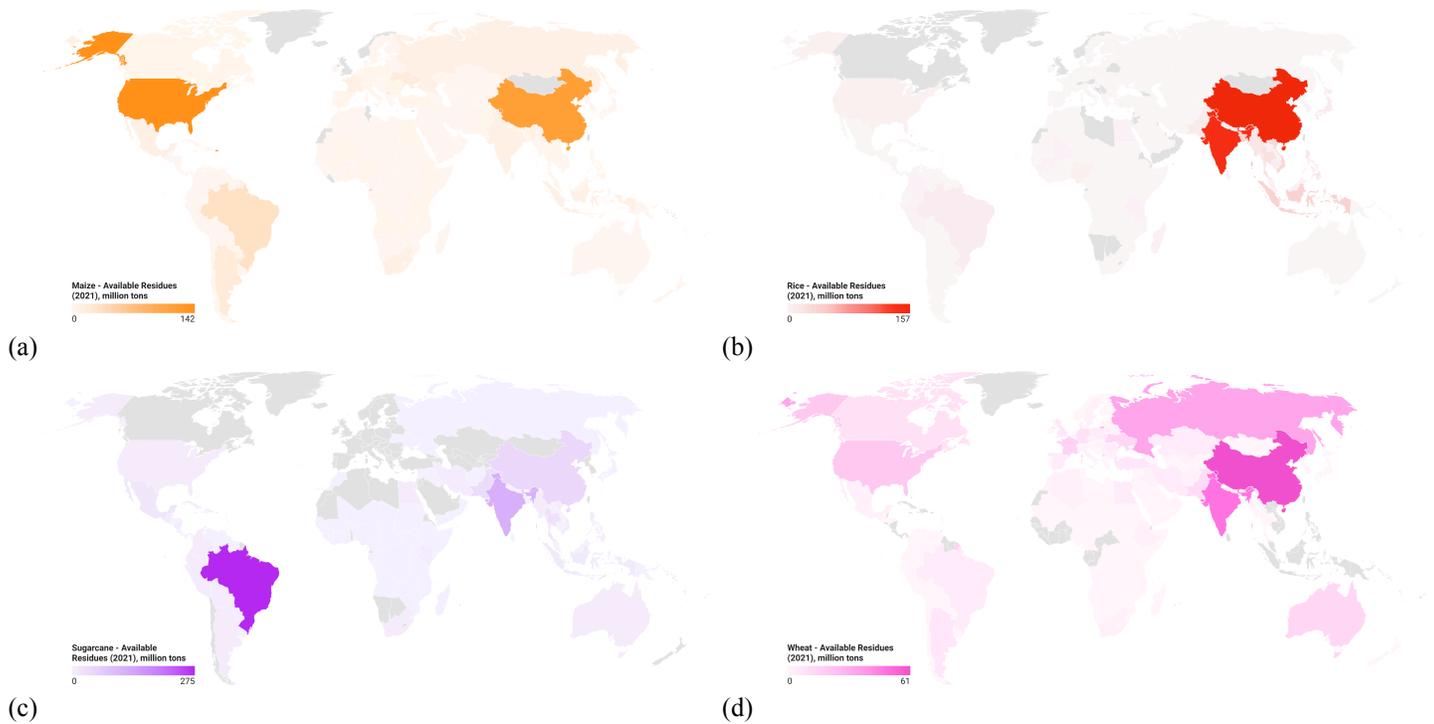

**Fig 4.** Available crop residues (2021), in million tons, interpolated linearly, for maize (a), rice (b), sugarcane (c), and wheat (d), by country

Bioenergy applications consumed an additional significant share of residues. Sugarcane bagasse-based bioenergy accounted for 223.0 million tons, with top utilizers being Brazil (71.3 million tons), India (56.3 million tons), China (15.6 million tons), Thailand (14.9 million tons), and Pakistan (8.7 million tons). For maize, rice, and wheat residues used in bioenergy, the total was 146.7 million tons, led by China (53.6 million tons), India (19.3 million tons), Brazil (12.1 million tons), Indonesia (9.5 million tons), and Thailand (5.0 million tons).

When aggregating both animal feed/bedding and bioenergy uses, the top five countries utilizing crop residues were Brazil (119.0 million tons), India (105.1 million tons), China (96.7 million tons), Pakistan (21.4 million tons), and the United States (21.1 million tons). Figure 5 highlights the 20 countries with the highest current maize, rice, sugarcane, and wheat crop residue use, for sugarcane bagasse for bioenergy; maize, rice, and wheat residues for bioenergy; and crop residues for animal feed and bedding. While a portion is already utilized, significant potential remains for expanding residue use in pellet production—especially in regions with high agricultural output and existing residue management infrastructure.

The final calculated volume of crop residues technically available for pelletization totals 1.4 billion tons, after accounting for sustainable removal rates, moisture content, and current competing uses. The countries with the highest available residue volumes are China (288.7 million tons), India (234.5 million tons), Brazil (221.5 million tons), United States (163.3 million tons), and Indonesia (51.0 million tons)—a reflection of their role as global agricultural powerhouses and major producers of the studied crops.

These available residues translate directly into potential energy production. By applying crop-specific lower heating values, the total energy content of the globally available residues can be estimated, representing a significant theoretical addition to renewable energy supply if converted into pellets. This amounts to 21.9 million TJ worldwide, with, accordingly, China (4.3 million TJ), Brazil (3.5 million TJ), India (3.4 million TJ), United States (2.6 million TJ), and Indonesia (0.7 million TJ) leading.

Figure 6 presents the global comparison of four key stages in this analysis: total crop production, total crop residues, final available crop residues, and the potential energy derived from these residues. Figure 7 further details the distribution of this potential energy by country, highlighting regional contributions and emphasizing the scale of untapped agricultural biomass resources worldwide.

### 4.3 Economic & Emissions Potential

By deriving country-specific capital expenditure (CAPEX) and operating expenditure (OPEX) based on adjusted price level indexes, this study established a foundation for modeling the financial viability of agricultural pellet plants across 179 countries.

On average, the capital expenditure required to establish an agricultural pellet plant with a production capacity of 40,080 tons per year over a 20-year period is approximately $5.3 million globally. Annual operating expenditures average $3.6 million, reflecting the combined costs of labor, raw materials, maintenance, and energy. Table S5 details the CAPEX and annual OPEX for each country, reflecting variations in labor, construction, raw materials, and energy costs. Based on these values and assuming a standardized production capacity of 40,080 tons per year over a 20-year operating period, the CLASP-P model calculated the minimum selling price required for a pellet plant in each country to break even under local cost conditions.

Globally, the minimum selling price of agricultural pellets—calculated using the CLASP-P model—was found to average $106 per ton. When normalized by energy content, the relevant metric, this equates to a levelized cost of energy (LCOE) of $6,600 per terajoule (TJ). This price sits notably higher than the global average cost of coal, estimated at $4,403 per TJ, yet remains competitive when compared to oil ($14,036 per TJ) and natural gas ($13,563 per TJ). These global averages, however,

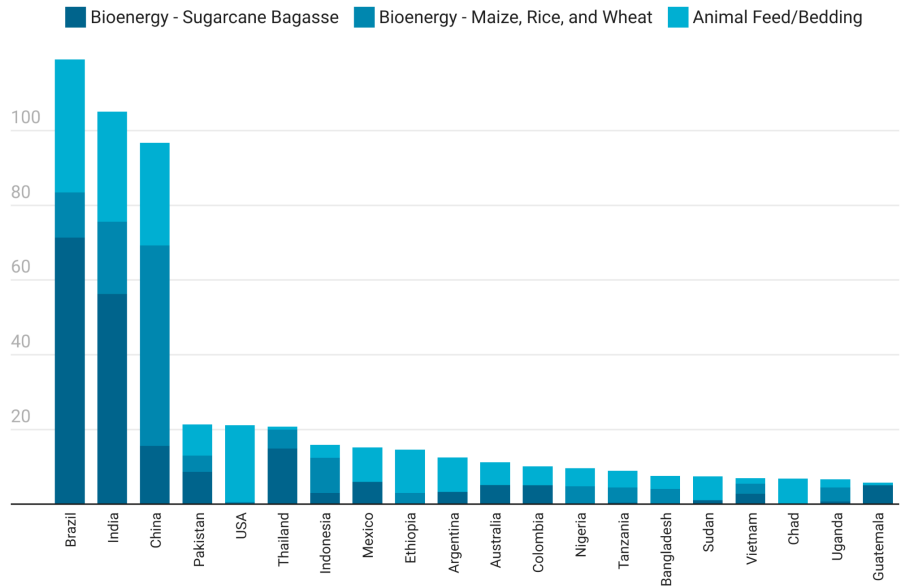

**Fig 5.** 20 countries with the highest current maize/rice/sugarcane/wheat crop residue use, for sugarcane bagasse bioenergy; maize, rice, and wheat residues for bioenergy; and crop residues for animal feed/bedding, in million tons.

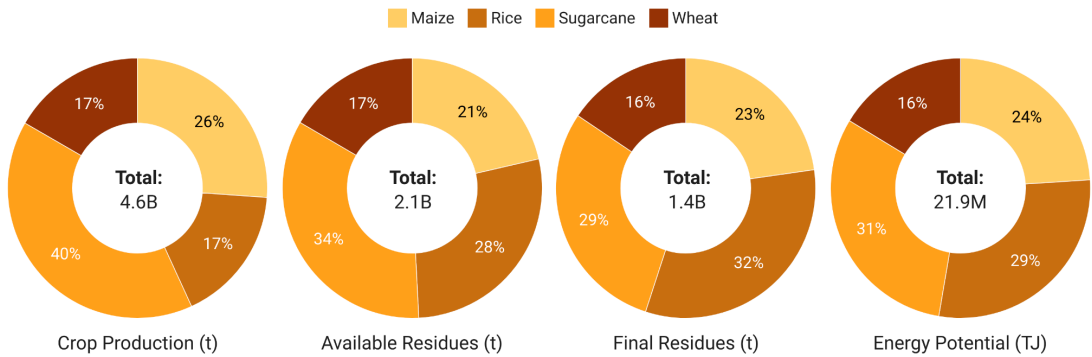

**Fig 6.** Breakdown of global crop production, available crop residues, final crop residues accounting for current use, and energy potential for maize, rice, sugarcane, and wheat.

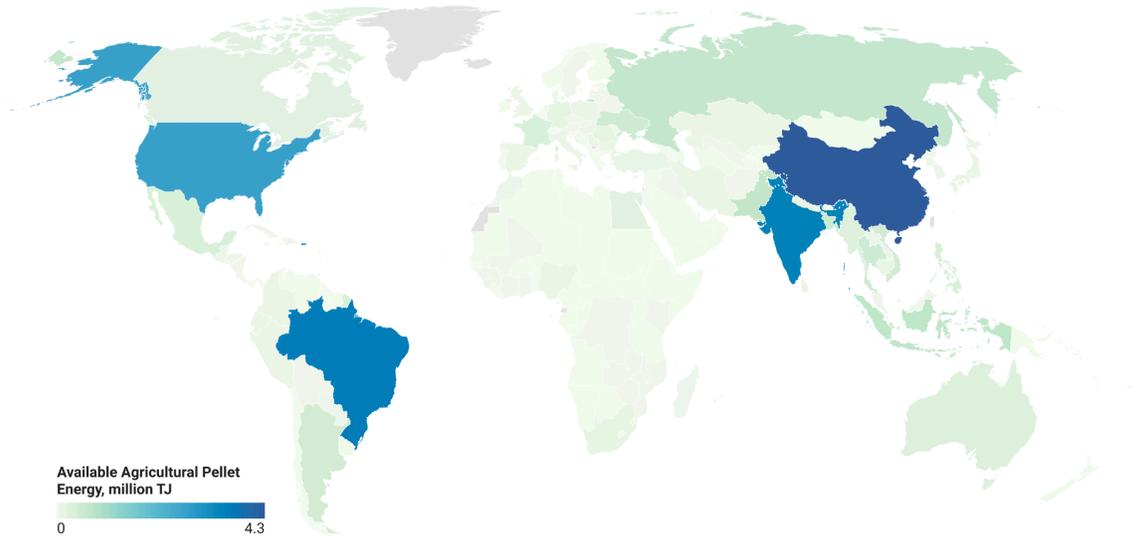

**Fig 7.** Available potential agricultural pellet energy, in million TJ, interpolated linearly, by country.

conceal significant regional variations, and country-specific pricing provides a stronger indication of economic attractiveness of pellet-based fossil fuel substitution, as well as which of the three studied fossil fuel should be replaced for the highest economic benefit. The model's output—the minimum selling price of pellets per ton—is presented in Underline Figure S1, alongside each country's average prices for coal, oil, and natural gas.

The integration of country-specific production costs and potential pellet energy quantities allows for a comprehensive assessment of the economic feasibility of pellet substitution for fossil fuels. Using the modeled pellet prices and potential energy, as well as the current consumption of fossil fuel energy [78] [79], this study quantified how much coal, oil, and natural gas energy could be realistically replaced by agricultural pellets in each country. Given that over 80% of global primary energy consumption is derived from coal, oil, and natural gas, this analysis positions agricultural pelletization within the context of mainstream energy markets. Figure 8 illustrates the global breakdown of energy consumption by fuel type, highlighting the critical role of these conventional energy sources—and, by extension, the significant opportunity for displacement.

In terms of current fossil fuel consumption, coal remains a dominant global energy source, with an estimated 161.6 million TJ consumed annually. The top five coal-consuming nations are China (87.6 million TJ), India (19.3 million TJ), United States of America (10.7 million TJ), Japan (4.9 million TJ), and South Africa (3.6 million TJ). This distribution underscores the heavy reliance on coal in both industrialized and emerging economies.

For oil, total global consumption reaches approximately 186.8 million TJ annually. The largest consumers are the United States of America (36.1 million TJ), China (29.4 million TJ), India (9.3 million TJ), Russian Federation (7.0 million TJ), and Japan (6.9 million TJ). Oil remains a key energy source across transport, industry, and power generation sectors worldwide.

Natural gas accounts for a global annual consumption of 145.9 million TJ. The leading consuming countries are the United States of America (30.6 million TJ), Russian Federation (15.5 million TJ), China (13.7 million TJ), Islamic Republic of Iran (8.3 million TJ), and Canada (4.3 million TJ). The widespread use of natural gas highlights its critical role in both energy generation and industrial applications globally.

To translate this theoretical potential in current consumption of fossil fuel energy and calculated pellet potential into actionable replacement plans, the RECOP model was employed for each country, optimizing the allocation of available pellet energy toward the most economically advantageous fossil fuel replacement. Prioritizing economic viability, Scenario A (Economically Optimized) was used to determine which fuels—coal, oil, or natural gas—should be replaced first based on cost savings. While this approach does not maximize emissions reductions, it reflects the commercial reality that economic incentives drive adoption decisions. RECOP dynamically ranks the three fossil fuels per country and allocates pellet energy accordingly, starting with the highest-ranked fuel and moving sequentially until all available pellet energy is assigned or all relevant fossil fuel demand is offset.

The impact of this optimized fuel replacement strategy is visualized in Figure 9, which plots each country's available pellet energy against its total energy consumption for coal, oil, and natural gas. The figure also displays the actual fossil fuel quantities replaced under RECOP's allocation, highlighting both the theoretical potential and the practical outcome of economically driven pellet substitution.

There exists significant variance in replacement potential across countries, influenced by factors such as energy market prices, residue availability, and existing fossil fuel dependence.

These results illustrate a clear global opportunity: agricultural residue pelletization can serve not only as a means of reducing emissions but as a commercially viable strategy for fossil fuel displacement. By leveraging cost-optimized deployment pathways, pelletization holds promise for enhancing energy security, promoting rural economic development, and contributing meaningfully to decarbonization efforts worldwide.

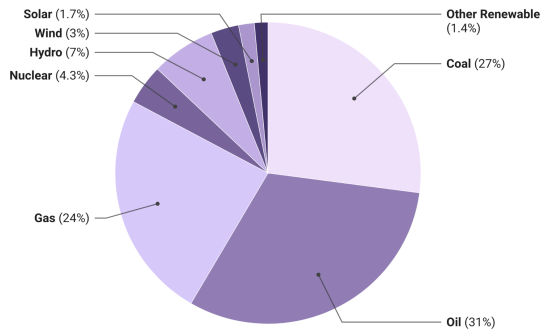

**Fig 8**. Breakdown of world energy consumption by generation, adapted from [78].

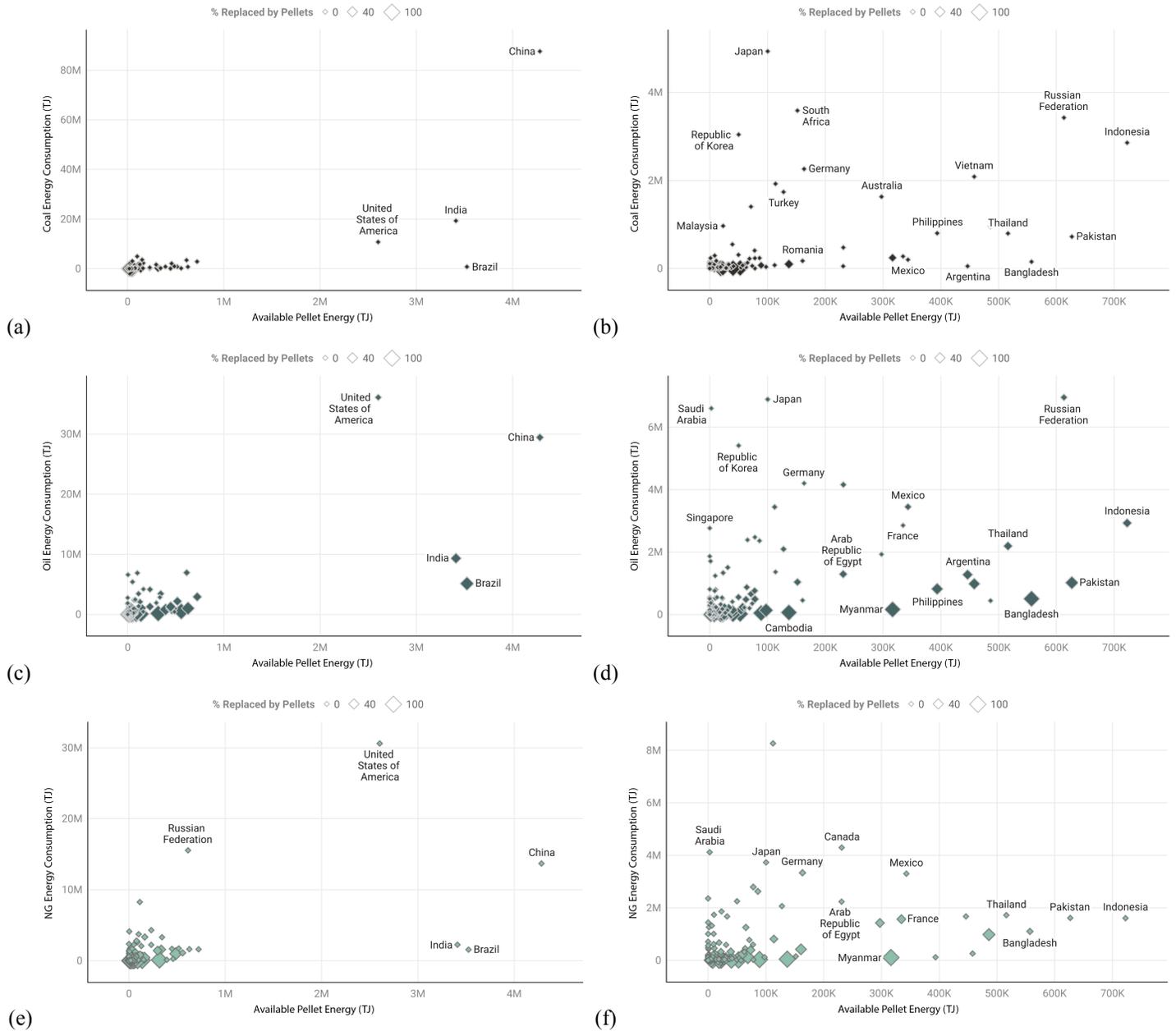

**Fig 9.** Available Pellet Energy in TJ and Energy Consumption in TJ for coal (a)(b), oil (c)(d), and natural gas (e)(f), by country, including (a)(c)(e) and excluding (b)(d)(f) Brazil, China, India, and the United States of America, and [(f) only] the Russian Federation. Symbol size reflects the percentage of corresponding fuel replaced by pellets, using values derived from RECOP.

### 4.3.1 Emissions Savings

The application of the RECOP model revealed clear patterns in the economic viability of fossil fuel replacement with agricultural pellets across countries. In 130 countries, oil emerged as the most economically beneficial fossil fuel to replace, reflecting its relatively high market price and widespread use in energy generation and industry. Conversely, natural gas ranked first in 48 countries, predominantly within Europe, where natural gas prices are comparatively higher due to market structures and supply dependencies. The ranking pattern was consistent across the model's prioritization algorithm: for countries where oil ranked first, natural gas ranked second, and vice versa. Notably, coal was consistently ranked third in all 178 countries analyzed, a result of its generally lower market price despite its significant emissions profile.

Globally, RECOP estimates that 4.5% of total coal, oil, and natural gas energy consumption could be replaced with agricultural pellets under the economically optimized scenario. While this may appear modest at a global scale, certain countries show remarkable replacement potential due to favorable local conditions such as residue availability, fuel prices, and market structure. The countries with the highest estimated replacement rates include Madagascar (100.0%), Malawi (100.0%), Sierra Leone (100.0%), Eswatini (97.3%), Belize (90.1%), Cambodia (65.4%), Myanmar (61.4%), Nepal (58.9%), Brazil (47.3%), and Guyana (40.8%). Figure 10 illustrates the percentage of fossil fuel energy replaced by agricultural pellets in each country under RECOP's economically optimized scenario and highlights which fossil fuel ranked as the most economically viable for substitution.

In aggregate, the modeled substitution of pellets for fossil fuels results in an estimated global emissions reduction of 1.3 billion tons of $CO_2$ equivalent yearly. The top contributors to this global reduction are China (263.1 million tons), Brazil (216.6 million tons), India (209.5 million tons), United States of America (159.9 million tons), and Indonesia (44.4 million tons). These figures reflect both the scale of fossil fuel use and the availability of agricultural residues in these countries, underscoring the intersection between agricultural production, energy policy, and emissions mitigation.

Figure 11(a) illustrates the emissions savings per year derived from RECOP, by country.

### 4.3.2 Economic Savings

Under the economically optimized scenario modeled by RECOP, the global economic savings potential from substituting agricultural pellets for fossil fuels is estimated at $163.1 billion per year. This figure reflects the combined impact of replacing higher-cost fossil fuels—primarily oil and natural gas—with competitively priced agricultural pellets across the countries analyzed. The largest projected economic savings are concentrated in key agricultural and industrial economies, with China ($49.7 billion), Brazil ($28.3 billion), India ($23.6 billion), United States of America ($19.9 billion), and Thailand ($5.5 billion) leading globally. These results emphasize the substantial financial benefit potential of agricultural pelletization when optimized for cost-effectiveness in national energy markets.

Figure 11(b) illustrates the economic savings per year derived from RECOP, by country.

### 4.3.3 Sensitivity Analysis

To assess the robustness of the economic savings projections, a sensitivity analysis was conducted to evaluate the impact of price fluctuations in both fossil fuels and agricultural pellets. Figure 12 presents the results of this analysis, mapping the variation in economic savings against changes in fuel and pellet prices.

Energy markets are historically volatile. Over the past decade, global coal and natural gas prices have experienced swings exceeding 200% within single years, while oil prices have shown comparable volatility with shifts of over 150%, according to long-term price data [65-67]. These price fluctuations have been driven by a combination of geopolitical tensions, supply chain disruptions, shifts in global demand, commodity market speculation, and changes in regulatory environments. Similar variability affects the key cost components used to calculate CAPEX and OPEX—namely, labor, electricity, raw materials, and construction costs—all of which are subject to local market dynamics and global economic pressures. This inherent volatility in both input costs and market prices underscores the importance of conducting comprehensive sensitivity analyses when evaluating the long-term economic viability of bioenergy projects like agricultural pelletization.

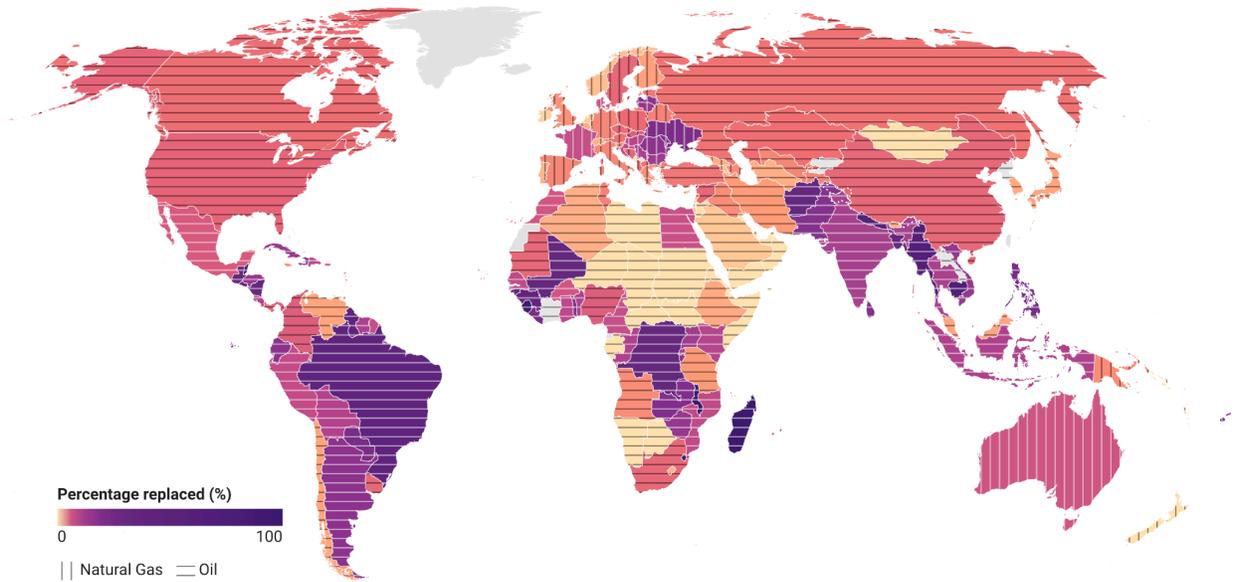

**Fig 10**. Percentage of coal, oil, and natural gas energy replaced by agricultural pellet energy, derived from RECOP, by country, interpolated in deciles. Vertical and horizontal lines indicate the fuel type considered most favorable for pellet replacement.

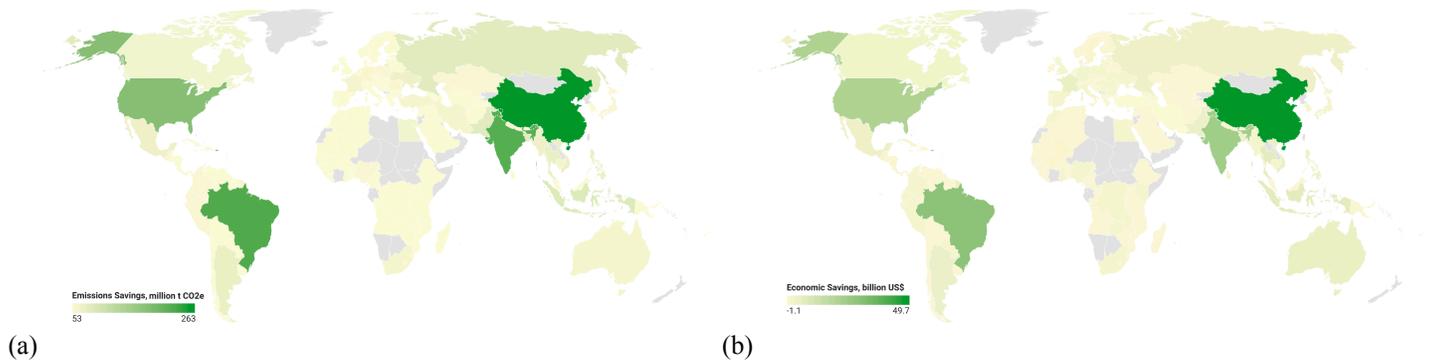

(a)                                                      (b)

**Fig 11.** Emissions (a) and economic (b) savings per year derived from RECOP, by country, interpolated linearly.

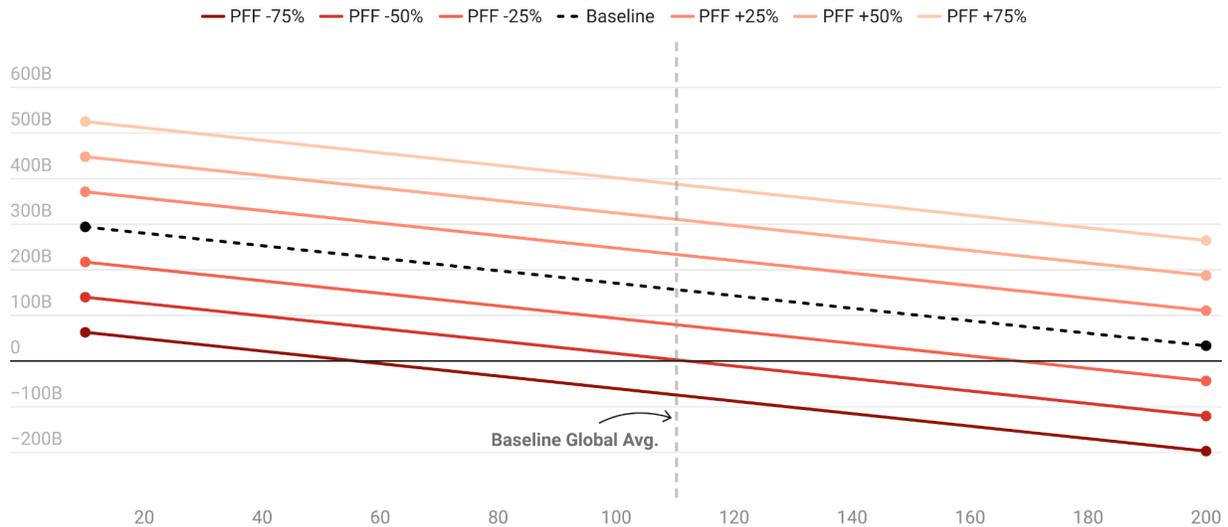

**Fig 12.** Sensitivity of global economic savings in US$ based on variable fossil fuel prices (PFF) and pellet prices from 10 to 200 $/t.

To capture these dynamics, the sensitivity analysis modeled seven fossil fuel price scenarios, varying from -75% to +75% in 25% increments, with the baseline price serving as a reference point. In parallel, eleven pellet price scenarios were modeled, ranging from $10 to $200, with the baseline global average being $106. This two-dimensional scenario matrix allowed for a detailed exploration of how different pricing environments could influence the relative economic attractiveness of pellet-based fossil fuel replacement strategies.

The sensitivity analysis revealed that the economic viability of agricultural pelletization remains robust under a wide range of price scenarios. Only when fossil fuel prices decreased by 75%, 50%, or 25% did RECOP return net negative economic savings, when pellet prices reached $60, $120, and $170 per ton, respectively. At the current global average estimated pellet price, only the most extreme -75% fossil fuel price scenario resulted in a negative economic outcome. Under baseline fossil fuel prices, economic savings ranged from $294.1 billion at a $10 pellet price to $33.7 billion at a $200 pellet price, representing a decline of nearly 90% but still maintaining positive savings across a wide pricing spectrum. These results reinforce that, even under volatile market conditions, agricultural pelletization holds strong potential as an economically viable alternative in global energy markets.

## 5. Discussion

### 5.1 Unlocking the Potential: Why Agricultural Pellets Matter

The global analysis conducted in this study underscores the transformative potential of agricultural residues as a strategic energy resource. With an estimated 1.44 billion tons of crop residues available worldwide after accounting for sustainable removal rates and competing uses, agricultural biomass represents a vast and largely untapped feedstock. This pool of material—concentrated in key agricultural economies such as China, India, Brazil, the United States, and Indonesia—provides a foundation for significant renewable energy deployment without compromising food production or land use.

Translating this biomass potential into energy terms, agricultural residues offer a theoretical capacity for displacing conventional fossil fuels at a meaningful scale. The integration of country-specific CAPEX, OPEX, and market price data through the CLASP-P model revealed that the minimum selling price for pellets remains competitive with several major fossil fuels, particularly oil and natural gas. Although the global average pellet price per terajoule exceeded that of coal, its alignment with oil and gas prices—and its favorable position in certain regional markets—highlights a substantial opportunity for targeted fuel substitution.

The RECOP model further demonstrated that agricultural pellets could replace 4.5% of current global coal, oil, and natural gas consumption, a figure that translates into both meaningful emissions reductions and considerable economic savings. The potential annual global emissions reduction of over 1.3 billion tons of $CO_2$ equivalent emphasizes the role agricultural pellets could play in decarbonizing national energy systems, particularly in leading economies such as China, Brazil, India, the United States, and Indonesia.

Economically, the estimated $163.1 billion in global annual savings positions agricultural pellets not just as an environmental solution, but as a commercially viable energy alternative. The countries with the highest projected savings—mirroring those with the largest available residues—underscore the symbiotic relationship between agricultural production, energy economics, and emissions mitigation.

Beyond macroeconomic gains, the strategic advantage of agricultural pellets lies in their decentralized potential. Unlike centralized energy systems dependent on large-scale fossil infrastructure, agricultural pelletization can leverage localized production networks, promoting rural economic development, reducing transportation costs, and enhancing energy security. The case study of a pelletization facility illustrated how this model can simultaneously deliver economic returns, reduce environmental pressures, and foster community participation.

## 5.2 Technical and Economic Challenges

While this study highlights the considerable economic and emissions potential of agricultural pellets, it also reveals a set of technical and economic challenges that could constrain large-scale adoption. These challenges are deeply linked to the nature of agricultural residues, the structure of biomass supply chains, and prevailing market dynamics.

One of the foremost technical barriers is the inherent variability and logistical complexity of agricultural residue collection. Unlike centralized fossil fuel extraction, agricultural residues are geographically dispersed and subject to seasonal production cycles, leading to inconsistent supply streams. This dispersion amplifies logistical challenges, including the cost and complexity of collection, transportation, and storage—factors that directly affect production costs and operational efficiency. The need for decentralized collection networks increases transportation distances and, consequently, energy input requirements, potentially undermining the carbon savings attributed to pellet use. These findings align with broader literature that identifies residue dispersal, high moisture content, and transportation difficulties as persistent obstacles in biomass utilization [5] [15] [32].

Economically, the capital and operating costs associated with agricultural pellet production remain significant. While this study found competitive minimum selling prices in certain markets, the overall viability is sensitive to plant scale, local labor and energy costs, and market access. Achieving economies of scale is critical; smaller-scale operations may face prohibitively high unit costs, limiting their ability to compete with established fossil fuels [18]. Additionally, the global pellet market is still benchmarked against wood pellets, which benefit from more mature supply chains and established trade networks. As such, agricultural pellets must match or undercut these benchmarks to gain meaningful market share [80].

The competitive dynamics are further complicated by fossil fuel pricing and carbon policy. In many markets, pellets cannot currently outcompete coal or petroleum coke on price alone without some form of policy intervention, such as carbon pricing, subsidies, or renewable energy incentives. Studies suggest that a carbon price of $30–50 per ton of $CO_2$ would be necessary for pellets to reach cost parity with coal in industrial and power sectors [6] [17]. Furthermore, governments have a critical role in shaping biomass markets—both by supporting technology development and by incentivizing supply chain formation through targeted policies [32].

On the technical side, advancements in pelletization technology remain essential, particularly in handling diverse agricultural residues with varying chemical compositions. Issues such as equipment wear, process optimization, and the technical challenges associated with densifying torrefied biomass continue to pose hurdles for operational efficiency and cost reduction [25].

In sum, while the global potential for agricultural pelletization is significant, unlocking this potential requires addressing multifaceted logistical, technical, and market challenges. These include improving collection and transportation systems, achieving production scale efficiencies, integrating with existing biomass supply chains, and ensuring supportive policy frameworks that enhance economic competitiveness. Without concerted efforts across these areas, the large-scale deployment of agricultural pellets may remain constrained despite their evident promise.

## 5.3 Policy Enablers: Incentives, Standards, and Investment

Unlocking the global potential of agricultural pelletization will likely depend not only on technological and economic factors but also on the presence of targeted policy enablers. Given the complexity and diversity of energy markets, agricultural systems, and regulatory environments worldwide, no single policy blueprint can universally apply. Instead, country-specific, regional, and even municipal policy frameworks must be calibrated to local circumstances, considering factors such as residue availability, market structure, energy demand, and existing industrial infrastructure.

Nevertheless, this study's findings suggest that certain policy mechanisms may serve as effective enablers in promoting agricultural pelletization. Financial incentives, such as subsidies for production, transportation, or infrastructure investment, can help offset high capital and operational costs—particularly in the early stages of market development. The experience of the studied pelletization facility, which benefits from local government subsidies as part of a rural industrial development strategy, exemplifies how targeted financial support can facilitate the viability of decentralized pellet operations. Such subsidies not only enhance economic feasibility but also foster rural employment and environmental management.

Beyond direct financial incentives, the establishment of clear technical standards for pellet production and quality assurance can foster market confidence and promote international trade. Standards ensure product consistency, facilitate integration into existing energy systems, and support the development of reliable supply chains.

In this regard, the wood pellet industry offers a valuable precedent, with its growth linked closely to the establishment of uniform specifications and certification schemes. Investment in research and development, particularly for improving pelletization technologies and optimizing supply chain logistics, remains another critical area for policy intervention. Public-private partnerships could accelerate innovation, especially in adapting equipment for diverse agricultural residues and improving the environmental performance of pellet production.

While this study underscores the promising global potential of agricultural pellets, it also highlights the importance of supportive policy frameworks in translating technical potential into market reality. Carefully designed incentives, standardization efforts, and strategic investments—tailored to local conditions—can play a pivotal role in scaling up agricultural pelletization as a viable component of the renewable energy transition.

## 5.4 Limitations of the Study

This study provides a comprehensive, data-driven evaluation of the global economic and emissions potential of agricultural residue pelletization. By combining empirical case study insights with quantitative modeling across 179 countries, it contributes valuable perspectives on an underexplored dimension of the renewable energy transition. The methodology integrates technical, economic, and policy considerations, offering an assessment of both opportunities and constraints.

However, as a global analysis, this study necessarily operates within the bounds of assumptions and generalized parameters that may not fully capture country-specific or region-specific realities. One core limitation stems from the use of globally harmonized datasets for crop production, energy prices, and residue availability, which may mask local variations in agricultural practices, residue collection logistics, and energy market dynamics.

Additionally, the study assumes that all technically available crop residues can be feasibly pelletized and used for energy within the same country—an assumption that overlooks significant logistical challenges. For instance, in countries with large geographic areas, agricultural production may be concentrated in one region while energy infrastructure is located in another, making the transport of bulky biomass economically impractical. This logistical disconnect could materially impact the real-world applicability of the modeled scenarios but is not accounted for within the current framework.

The settings on the RECOP model also introduces important analytical boundaries. This study prioritized the economically optimized scenario to reflect market-driven decision-making. While this approach identified a potential 1.3 billion tons of $CO_2$ equivalent emissions savings and \$163.1 billion in global economic savings, it does not capture the maximum theoretical emissions reduction possible. Under a purely emissions-optimized scenario, the potential savings from RECOP rise to 2.2 billion tons of $CO_2$ equivalent—an increase of approximately 60%—but at the cost of generating an overall \$54.7 billion economic loss compared to the economically optimized model, over $200 billion lower than this study's scenario. This trade-off highlights the tension between environmental objectives and market viability, a nuance that the model's single-scenario outputs may oversimplify.

Further limitations include assumptions related to energy price stability, carbon market mechanisms, and technological performance. This study's sensitivity analysis partially addressed price volatility, yet it cannot fully capture future market fluctuations, policy changes, or technological innovations that may alter the competitive landscape for agricultural pellets.

Given these constraints, further research is recommended to refine and localize the findings of this study. Future analyses could incorporate region-specific supply chain assessments, create country-wide geographical optimizations for pellet transportation, and integrate dynamic energy market modeling. In particular, in-depth case studies across different geographic contexts could offer critical insights into the practical challenges and enablers of scaling agricultural pelletization within diverse national energy systems.

## 6. Conclusions

This study set out to quantify the global economic and emissions reduction potential of agricultural residue pelletization, combining empirical insights from the case study with large-scale data modeling across 179 countries. The findings underscore the viability of agricultural pellets as a strategic complement to global decarbonization efforts—offering both environmental benefits and economic opportunities.

With an estimated **1.44 billion tons** of crop residues available for sustainable pelletization, the theoretical energy potential of agricultural residues is significant. When translated into market terms, the minimum selling price of pellets calculated through the CLASP-P model demonstrated competitiveness in many national contexts, especially when benchmarked against oil and natural gas. Furthermore, the RECOP model revealed a **4.5% replacement potential** of fossil fuel consumption, amounting to a global emissions reduction of over **1.3 billion tons** of $CO_2$ equivalent and an annual economic savings exceeding **$163 billion** under economically optimized conditions.

This global potential resonates with the dual imperative highlighted at the outset of this study: the need for renewable energy solutions that are both economically viable and environmentally sustainable. Agricultural pellets emerge as a response to both challenges—leveraging underutilized agricultural residues to produce a renewable energy source that aligns with global decarbonization goals while also fostering economic returns. The combination of scalable residue availability, proven technical feasibility, and demonstrated market competitiveness suggests that agricultural pelletization is not merely a theoretical opportunity, but a practical strategy for achieving tangible energy and climate outcomes.

Ultimately, the results position agricultural pellets as a critical component of the renewable energy transition. By leveraging sustainably harvested residues, advancing cost-competitive production models, and applying economically optimized deployment strategies, agricultural pelletization presents a realistic and scalable pathway for countries to align climate action with economic development.

However, realizing this potential will require addressing operational challenges, refining supportive policies, and fostering further research to localize global models and integrate region-specific realities. With targeted investment and policy support, agricultural pellets could become a cornerstone of sustainable energy systems worldwide.

## Declaration of Competing Interest

The authors declare that they have no known competing financial interests or personal relationships that could have appeared to influence the work reported in this paper.

## Data Availability

Data used for the study will be made available upon request.

**Supplementary Information Section**

This supplementary information section provides a summary of additional data/information that informs this study's model development and analysis.

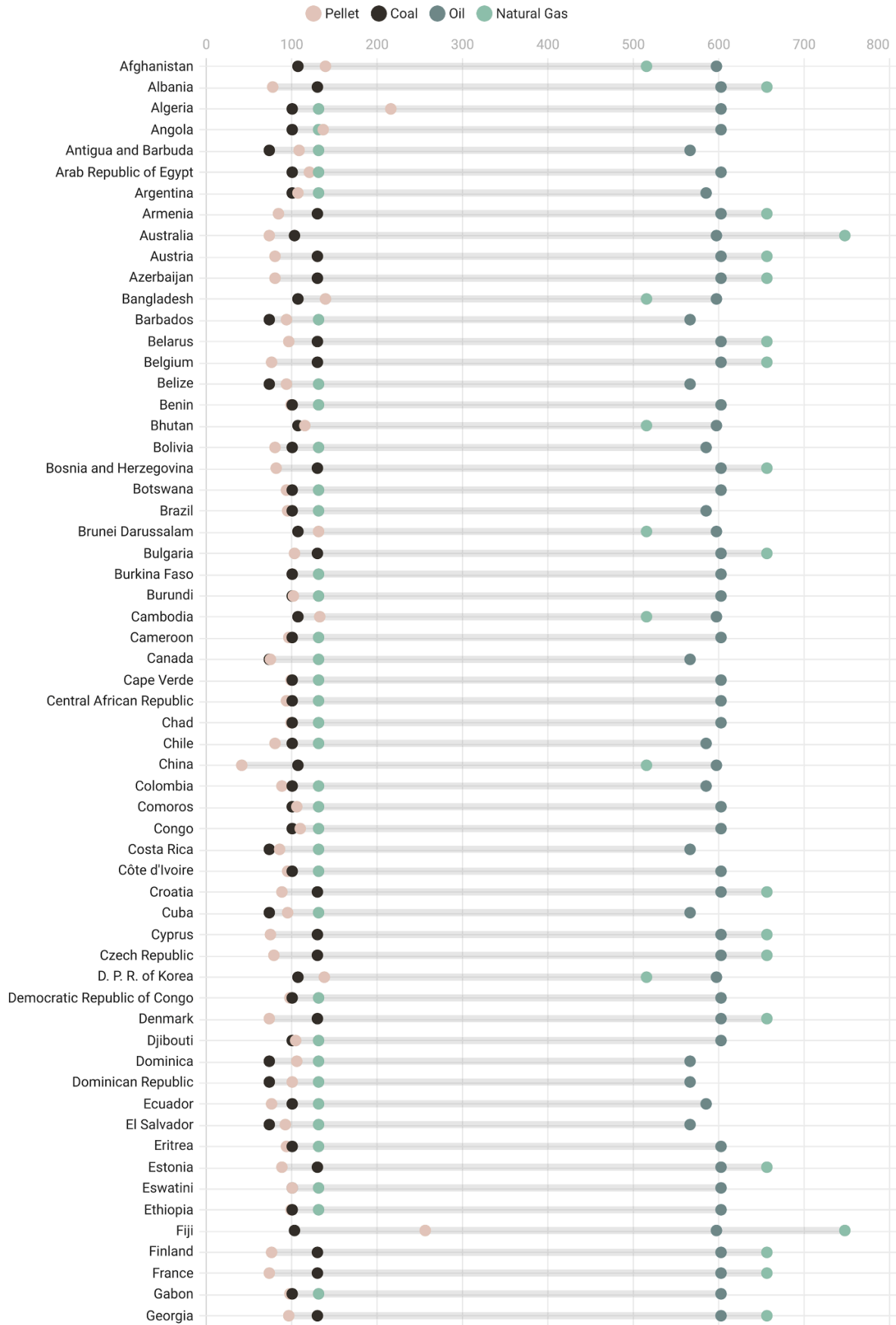

**Fig S1.** Country-specific prices for coal, crude oil, natural gas, and price for agricultural pellets derived from CLASP-P, in US$/t.

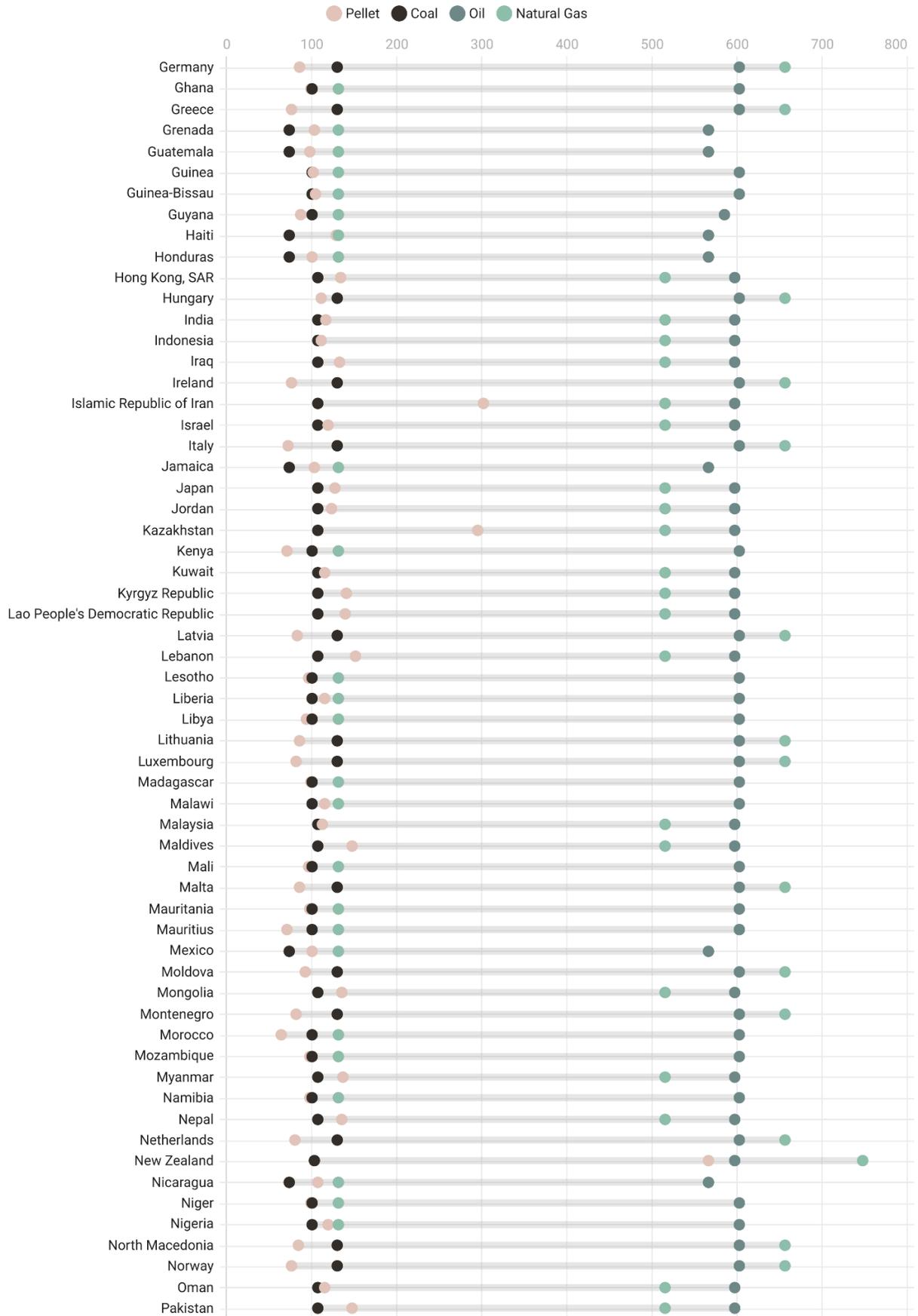

**Fig S1.** (Continued)

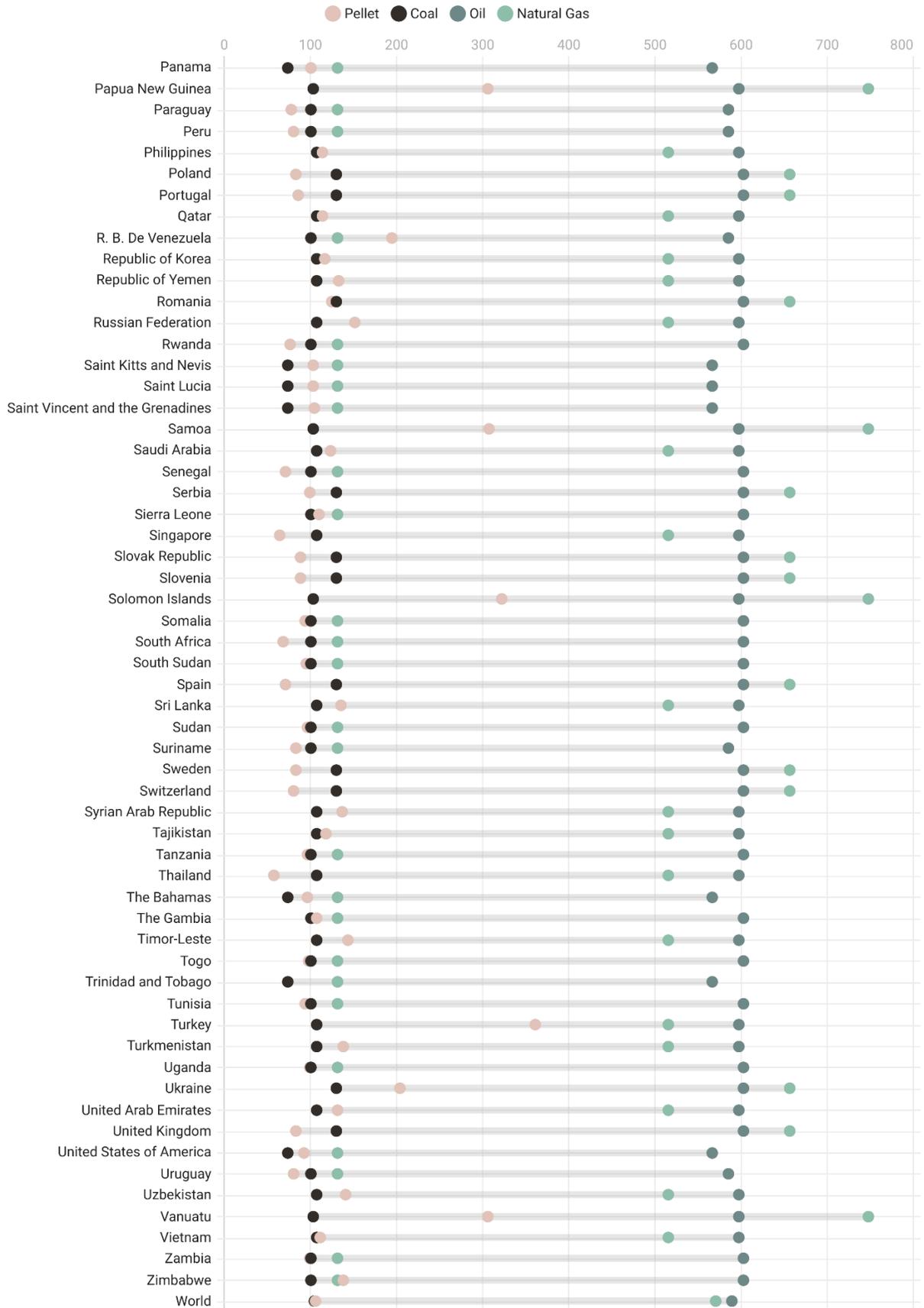

**Fig S1.** (Continued)

| Country | Maize | Rice | Sugarcane | Wheat | Total |
|---|---|---|---|---|---|
| Afghanistan | 0.18 | 0.46 | 0.04 | 3.90 | 4.58 |
| Albania | 0.41 | 0.00 | - | 0.23 | 0.64 |
| Algeria | 0.02 | 0.00 | - | 2.17 | 2.19 |
| Angola | 2.97 | 0.01 | 0.95 | 0.00 | 3.93 |
| Antigua and Barbuda | 0.00 | - | 0.00 | - | 0.00 |
| Arab Republic of Egypt | 7.50 | 4.84 | 12.36 | 9.00 | 33.70 |
| Argentina | 60.53 | 1.45 | 18.63 | 17.64 | 98.25 |
| Armenia | 0.01 | 0.00 | - | 0.10 | 0.10 |
| Australia | 0.31 | 0.42 | 31.13 | 31.92 | 63.78 |
| Austria | 2.43 | - | - | 1.55 | 3.98 |
| Azerbaijan | 0.28 | 0.01 | - | 1.84 | 2.13 |
| Bangladesh | 4.12 | 56.94 | 3.33 | 1.09 | 65.48 |
| Barbados | 0.00 | - | 0.09 | - | 0.09 |
| Belarus | 1.15 | - | - | 2.44 | 3.59 |
| Belgium | 0.45 | 0.00 | - | 1.63 | 2.08 |
| Belize | 0.11 | 0.01 | 1.89 | - | 2.02 |
| Benin | 1.63 | 0.52 | 0.08 | - | 2.23 |
| Bhutan | 0.03 | 0.04 | 0.00 | 0.00 | 0.07 |
| Bolivia | 1.22 | 0.55 | 10.09 | 0.34 | 12.20 |
| Bosnia and Herzegovina | 0.89 | - | - | 0.31 | 1.21 |
| Botswana | 0.07 | - | - | 0.00 | 0.07 |
| Brazil | 88.46 | 11.66 | 715.66 | 7.87 | 823.66 |
| Brunei Darussalam | - | 0.00 | - | - | 0.00 |
| Bulgaria | 3.43 | 0.06 | - | 7.34 | 10.83 |
| Burkina Faso | 1.91 | 0.45 | 0.51 | - | 2.87 |
| Burundi | 0.28 | 0.12 | 0.20 | 0.01 | 0.61 |
| Cambodia | 0.92 | 11.41 | 2.99 | - | 15.32 |
| Cameroon | 2.10 | 0.36 | 1.23 | 0.00 | 3.69 |
| Canada | 13.98 | - | - | 22.30 | 36.28 |
| Cape Verde | 0.00 | - | 0.02 | - | 0.02 |
| Central African Republic | 0.09 | 0.01 | 0.13 | - | 0.23 |
| Chad | 0.36 | 0.24 | 0.41 | 0.00 | 1.02 |
| Chile | 0.79 | 0.15 | - | 1.35 | 2.29 |
| China | 272.76 | 214.40 | 107.26 | 136.95 | 731.38 |
| Colombia | 1.59 | 3.33 | 24.03 | 0.01 | 28.96 |
| Comoros | 0.01 | 0.03 | - | - | 0.04 |
| Congo | 0.01 | 0.00 | 0.72 | - | 0.73 |
| Costa Rica | 0.01 | 0.15 | 4.30 | - | 4.45 |
| Côte d'Ivoire | 1.14 | 1.66 | 2.10 | - | 4.90 |
| Croatia | 2.24 | 0.00 | - | 0.99 | 3.23 |
| Cuba | 0.24 | 0.23 | 11.21 | - | 11.67 |

| Country | Maize | Rice | Sugarcane | Wheat | Total |
|---|---|---|---|---|---|
| Cyprus | - | 0.00 | - | 0.03 | 0.03 |
| Czech Republic | 0.99 | 0.00 | - | 4.96 | 5.95 |
| D. P. R. of Korea | 2.30 | 1.85 | - | 0.09 | 4.24 |
| Democratic Republic of Congo | 2.24 | 1.58 | 2.18 | 0.01 | 6.01 |
| Denmark | 0.05 | 0.00 | - | 4.05 | 4.09 |
| Djibouti | 0.00 | - | 0.00 | - | 0.00 |
| Dominica | 0.00 | - | 0.00 | - | 0.01 |
| Dominican Republic | 0.05 | 1.01 | 5.51 | - | 6.57 |
| Ecuador | 1.70 | 1.50 | 11.37 | 0.01 | 14.59 |
| El Salvador | 0.88 | 0.02 | 7.51 | - | 8.41 |
| Eritrea | 0.02 | - | - | 0.03 | 0.05 |
| Estonia | 0.00 | 0.00 | - | 0.74 | 0.74 |
| Eswatini | 0.10 | 0.00 | 5.72 | 0.00 | 5.82 |
| Ethiopia | 10.72 | 0.20 | 1.17 | 5.21 | 17.31 |
| Fiji | 0.00 | 0.01 | 1.42 | - | 1.43 |
| Finland | 0.00 | 0.00 | - | 0.69 | 0.69 |
| France | 15.36 | 0.06 | - | 36.56 | 51.98 |
| Gabon | 0.05 | 0.00 | 0.29 | - | 0.34 |
| Georgia | 0.23 | - | - | 0.14 | 0.37 |
| Germany | 4.46 | 0.00 | - | 21.46 | 25.92 |
| Ghana | 3.50 | 1.23 | 0.15 | - | 4.89 |
| Greece | 1.32 | 0.24 | - | 1.06 | 2.62 |
| Grenada | 0.00 | - | 0.01 | - | 0.01 |
| Guatemala | 1.96 | 0.03 | 27.76 | 0.00 | 29.75 |
| Guinea | 0.80 | 2.48 | 0.32 | - | 3.59 |
| Guinea-Bissau | 0.02 | 0.21 | 0.01 | - | 0.24 |
| Guyana | 0.00 | 0.56 | 1.28 | - | 1.84 |
| Haiti | 0.20 | 0.16 | 1.47 | - | 1.82 |
| Honduras | 0.67 | 0.05 | 4.76 | 0.00 | 5.49 |
| Hong Kong, SAR | - | 0.00 | - | - | 0.00 |
| Hungary | 6.42 | 0.01 | - | 5.29 | 11.72 |
| India | 31.65 | 195.43 | 405.40 | 109.59 | 742.06 |
| Indonesia | 20.01 | 54.42 | 32.20 | - | 106.63 |
| Iraq | 0.37 | 0.42 | 0.00 | 4.23 | 5.03 |
| Ireland | 0.00 | 0.00 | - | 0.63 | 0.63 |
| Islamic Republic of Iran | 0.32 | 1.60 | 8.26 | 10.09 | 20.27 |
| Israel | 0.07 | - | - | 0.15 | 0.22 |
| Italy | 6.08 | 1.46 | - | 7.29 | 14.83 |
| Jamaica | 0.00 | 0.00 | 0.50 | - | 0.50 |
| Japan | 0.00 | 10.53 | 1.31 | 1.10 | 12.93 |
| Jordan | 0.03 | - | - | 0.03 | 0.06 |

| Country | Maize | Rice | Sugarcane | Wheat | Total |
|---|---|---|---|---|---|
| Kazakhstan | 1.13 | 0.50 | - | 11.81 | 13.45 |
| Kenya | 3.30 | 0.19 | 7.78 | 0.25 | 11.52 |
| Kuwait | 0.01 | - | - | 0.00 | 0.01 |
| Kyrgyz Republic | 0.69 | 0.05 | - | 0.36 | 1.10 |
| Lao People's Democratic Republic | 1.05 | 3.87 | 1.89 | - | 6.81 |
| Latvia | 0.00 | 0.00 | - | 2.41 | 2.41 |
| Lebanon | 0.00 | - | 0.00 | 0.10 | 0.10 |
| Lesotho | 0.09 | - | - | 0.01 | 0.10 |
| Liberia | - | 0.26 | 0.28 | | 0.53 |
| Libya | 0.00 | - | - | 0.13 | 0.13 |
| Lithuania | 0.10 | 0.00 | - | 4.25 | 4.35 |
| Luxembourg | 0.00 | 0.00 | - | 0.08 | 0.08 |
| Madagascar | 0.23 | 4.39 | 3.12 | 0.00 | 7.75 |
| Malawi | 4.58 | 0.15 | 3.16 | 0.00 | 7.89 |
| Malaysia | 0.07 | 2.42 | 0.02 | - | 2.52 |
| Maldives | 0.00 | - | - | - | 0.00 |
| Mali | 3.60 | 2.42 | 0.65 | 0.02 | 6.69 |
| Malta | 0.00 | 0.00 | - | 0.00 | 0.00 |
| Mauritania | 0.01 | 0.43 | - | 0.01 | 0.45 |
| Mauritius | 0.00 | 0.00 | 2.67 | - | 2.67 |
| Mexico | 27.50 | 0.26 | 55.49 | 3.28 | 86.53 |
| Moldova | 2.79 | 0.00 | - | 1.57 | 4.36 |
| Mongolia | - | - | - | 0.57 | 0.57 |
| Montenegro | 0.00 | - | - | 0.00 | 0.00 |
| Morocco | 0.05 | 0.05 | 0.61 | 7.54 | 8.26 |
| Mozambique | 2.10 | 0.19 | 2.99 | 0.02 | 5.29 |
| Myanmar | 2.30 | 24.91 | 11.65 | 0.10 | 38.96 |
| Namibia | 0.09 | - | - | 0.02 | 0.11 |
| Nepal | 3.00 | 5.62 | 3.18 | 2.13 | 13.93 |
| Netherlands | 0.17 | 0.00 | - | 0.95 | 1.12 |
| New Zealand | 0.21 | 0.00 | - | 0.42 | 0.63 |
| Nicaragua | 0.38 | 0.49 | 7.11 | - | 7.97 |
| Niger | 0.01 | 0.02 | 0.45 | 0.00 | 0.48 |
| Nigeria | 12.75 | 8.34 | 1.50 | 0.09 | 22.68 |
| North Macedonia | 0.13 | 0.02 | - | 0.24 | 0.40 |
| Norway | - | - | - | 0.27 | 0.27 |
| Oman | 0.02 | - | 0.00 | 0.00 | 0.02 |
| Pakistan | 10.63 | 13.98 | 88.65 | 27.46 | 140.73 |
| Panama | 0.14 | 0.41 | 2.41 | - | 2.96 |
| Papua New Guinea | 0.01 | 0.00 | 0.35 | - | 0.37 |
| Paraguay | 4.09 | 1.18 | 7.22 | 0.93 | 13.42 |

| Country | Maize | Rice | Sugarcane | Wheat | Total |
|---|---|---|---|---|---|
| Peru | 1.58 | 3.47 | 9.83 | 0.20 | 15.09 |
| Philippines | 8.30 | 19.96 | 26.28 | - | 54.54 |
| Poland | 7.32 | 0.00 | - | 11.89 | 19.22 |
| Portugal | 0.75 | 0.18 | - | 0.07 | 1.00 |
| Qatar | 0.00 | - | - | 0.00 | 0.00 |
| R. B. De Venezuela | 1.54 | 0.79 | 3.21 | 0.00 | 5.55 |
| Republic of Korea | 0.08 | 5.21 | - | 0.03 | 5.32 |
| Republic of Yemen | 0.04 | - | 0.00 | 0.13 | 0.16 |
| Romania | 14.82 | 0.01 | - | 10.43 | 25.27 |
| Russian Federation | 15.24 | 1.08 | 0.00 | 76.06 | 92.37 |
| Rwanda | 0.48 | 0.13 | 0.10 | 0.01 | 0.73 |
| Saint Kitts and Nevis | - | - | 0.00 | - | 0.00 |
| Saint Lucia | 0.00 | - | 0.00 | - | 0.00 |
| Saint Vincent and the Grenadines | 0.00 | 0.00 | - | - | 0.00 |
| Samoa | - | - | 0.00 | - | 0.00 |
| Saudi Arabia | 0.06 | 0.00 | - | 0.61 | 0.67 |
| Senegal | 0.75 | 1.38 | 1.39 | - | 3.52 |
| Serbia | 6.03 | - | - | 3.44 | 9.47 |
| Sierra Leone | 0.02 | 1.98 | 0.08 | - | 2.08 |
| Singapore | - | - | 0.00 | - | 0.00 |
| Slovak Republic | 1.58 | 0.00 | - | 2.00 | 3.58 |
| Slovenia | 0.39 | 0.00 | - | 0.15 | 0.54 |
| Solomon Islands | - | 0.00 | - | - | 0.00 |
| Somalia | 0.08 | 0.00 | 0.21 | 0.00 | 0.29 |
| South Africa | 16.87 | 0.00 | 17.99 | 2.26 | 37.12 |
| South Sudan | 0.18 | 0.03 | 0.00 | 0.00 | 0.20 |
| Spain | 4.60 | 0.62 | - | 8.56 | 13.78 |
| Sri Lanka | 0.47 | 5.15 | 0.83 | - | 6.45 |
| Sudan | 0.02 | 0.03 | 5.33 | 0.60 | 5.98 |
| Suriname | 0.00 | 0.26 | 0.09 | - | 0.35 |
| Sweden | 0.01 | 0.00 | - | 3.03 | 3.04 |
| Switzerland | 0.10 | - | - | 0.40 | 0.50 |
| Syrian Arab Republic | 0.31 | 0.00 | 0.00 | 1.95 | 2.26 |
| Tajikistan | 0.24 | 0.07 | - | 0.85 | 1.16 |
| Tanzania | 7.04 | 2.69 | 3.52 | 0.07 | 13.31 |
| Thailand | 5.30 | 33.58 | 66.28 | 0.00 | 105.16 |
| The Bahamas | 0.00 | - | 0.06 | - | 0.06 |
| The Gambia | 0.02 | 0.04 | - | - | 0.06 |
| Timor-Leste | 0.06 | 0.05 | - | - | 0.10 |
| Togo | 0.93 | 0.15 | - | - | 1.08 |
| Trinidad and Tobago | 0.01 | 0.00 | 0.00 | - | 0.01 |

| Country | Maize | Rice | Sugarcane | Wheat | Total |
|---|---|---|---|---|---|
| Tunisia | - | - | - | 1.19 | 1.19 |
| Turkey | 6.75 | 1.00 | - | 17.65 | 25.40 |
| Turkmenistan | 0.01 | 0.08 | - | 1.37 | 1.46 |
| Uganda | 2.80 | 0.30 | 5.37 | 0.03 | 8.50 |
| Ukraine | 42.11 | 0.05 | 0.00 | 32.18 | 74.34 |
| United Arab Emirates | 0.02 | - | - | 0.00 | 0.02 |
| United Kingdom | - | - | - | 13.99 | 13.99 |
| United States of America | 383.94 | 8.70 | 29.96 | 44.79 | 467.40 |
| Uruguay | 0.77 | 1.31 | 0.46 | 0.94 | 3.47 |
| Uzbekistan | 0.59 | 0.33 | - | 5.98 | 6.91 |
| Vanuatu | 0.00 | - | - | - | 0.00 |
| Vietnam | 4.45 | 43.85 | 10.74 | - | 59.04 |
| Zambia | 3.62 | 0.07 | 5.10 | 0.21 | 8.99 |
| Zimbabwe | 1.47 | 0.00 | 3.45 | 0.34 | 5.26 |
| World | 1210.24 | 787.29 | 1859.39 | 770.88 | 4627.80 |

**Table S1.** Crop production for maize, rice, sugarcane, and wheat, by country, adapted from [34-37], in million tons. Color scale indicates relative production volume, with green shades representing higher production values.

| Country | Maize | Rice | Sugarcane | Wheat |
|---|---|---|---|---|
| Austria | 73.74 | 87.74 | - | 86.27 |
| Belgium | 73.74 | 87.74 | - | 86.27 |
| Bulgaria | 73.74 | 87.74 | - | 86.27 |
| Cambodia | 85.13 | 86.99 | 46.57 | 86.27 |
| Canada | 73.74 | - | - | 86.27 |
| China | 85.13 | 86.99 | 46.57 | 86.27 |
| Costa Rica | 73.74 | - | - | - |
| Croatia | 73.74 | 87.74 | - | 86.27 |
| Cyprus | 73.74 | 87.74 | - | 86.27 |
| Czech Republic | 73.74 | 87.74 | - | 86.27 |
| D. P. R. of Korea | 85.13 | 86.99 | 46.57 | 86.27 |
| Denmark | 73.74 | 87.74 | - | 86.27 |
| Estonia | 73.74 | 87.74 | - | 86.27 |
| Finland | 73.74 | 87.74 | - | 86.27 |
| France | 73.74 | 87.74 | - | 86.27 |
| Greece | 73.74 | 87.74 | - | 86.27 |
| Hungary | 73.74 | 87.74 | - | 86.27 |
| India | 56.67 | 90.46 | 34.47 | - |
| Indonesia | 85.13 | 86.99 | 46.57 | 86.27 |
| Ireland | 73.74 | 87.74 | - | 86.27 |
| Italy | 73.74 | 87.74 | - | 86.27 |
| Japan | 85.13 | 86.99 | 46.57 | 86.27 |
| Lao People's Democratic Republic | 85.13 | 86.99 | 46.57 | 86.27 |
| Latvia | 73.74 | 87.74 | - | 86.27 |
| Lithuania | 73.74 | 87.74 | - | 86.27 |
| Luxembourg | 73.74 | 87.74 | - | 86.27 |
| Malaysia | 85.13 | 86.99 | 46.57 | 86.27 |
| Malta | 73.74 | 87.74 | - | 86.27 |
| Mongolia | 85.13 | 86.99 | 46.57 | 86.27 |
| Myanmar | 85.13 | 86.99 | 46.57 | 86.27 |
| Netherlands | 73.74 | 87.74 | - | 86.27 |
| Philippines | 85.13 | 86.99 | 46.57 | 86.27 |
| Poland | 73.74 | 87.74 | - | 86.27 |
| Portugal | 73.74 | 87.74 | - | 86.27 |
| Republic of Korea | 85.13 | 86.99 | 46.57 | 86.27 |
| Romania | 73.74 | 87.74 | - | 86.27 |
| Singapore | 85.13 | 86.99 | 46.57 | 86.27 |
| Slovak Republic | 73.74 | 87.74 | - | 86.27 |
| Slovenia | 73.74 | 87.74 | - | 86.27 |

| Country | Maize | Rice | Sugarcane | Wheat |
|---|---|---|---|---|
| Spain | 73.74 | 87.74 | - | 86.27 |
| Sweden | 73.74 | 87.74 | - | 86.27 |
| Thailand | 85.13 | 86.99 | 46.57 | 86.27 |
| Timor-Leste | 85.13 | 86.99 | 46.57 | 86.27 |
| United Kingdom | 73.74 | 87.74 | - | 86.27 |
| Vietnam | 85.13 | 86.99 | 46.57 | 86.27 |
| World Average | 73.74 | 87.74 | 43.88 | 86.27 |

**Table S2.** Dry matter as a percentage of fresh weight of crop residues, by country, adapted from [42].

| Country | Maize | Rice | Sugarcane | Wheat | Total |
|---|---|---|---|---|---|
| Afghanistan | 0.18 | 0.64 | 0.04 | 5.07 | 5.93 |
| Albania | 0.41 | 0.00 | - | 0.29 | 0.71 |
| Algeria | 0.02 | 0.00 | - | 2.82 | 2.84 |
| Angola | 2.97 | 0.01 | 0.95 | 0.00 | 3.94 |
| Antigua and Barbuda | 0.00 | - | 0.00 | - | 0.00 |
| Arab Republic of Egypt | 7.50 | 6.78 | 12.36 | 11.70 | 38.34 |
| Argentina | 60.53 | 2.03 | 18.63 | 22.94 | 104.13 |
| Armenia | 0.01 | 0.00 | - | 0.13 | 0.13 |
| Australia | 0.31 | 0.59 | 31.13 | 41.50 | 73.53 |
| Austria | 2.43 | - | - | 2.01 | 4.45 |
| Azerbaijan | 0.28 | 0.01 | - | 2.39 | 2.68 |
| Bangladesh | 4.12 | 79.72 | 3.33 | 1.41 | 88.58 |
| Barbados | 0.00 | - | 0.09 | - | 0.09 |
| Belarus | 1.15 | - | - | 3.17 | 4.32 |
| Belgium | 0.45 | 0.00 | - | 2.12 | 2.57 |
| Belize | 0.11 | 0.02 | 1.89 | - | 2.02 |
| Benin | 1.63 | 0.73 | 0.08 | - | 2.44 |
| Bhutan | 0.03 | 0.06 | 0.00 | 0.00 | 0.09 |
| Bolivia | 1.22 | 0.77 | 10.09 | 0.44 | 12.52 |
| Bosnia and Herzegovina | 0.89 | - | - | 0.41 | 1.30 |
| Botswana | 0.07 | - | - | 0.00 | 0.07 |
| Brazil | 88.46 | 16.32 | 715.66 | 10.24 | 830.68 |
| Brunei Darussalam | - | 0.00 | - | - | 0.00 |
| Bulgaria | 3.43 | 0.08 | - | 9.55 | 13.05 |
| Burkina Faso | 1.91 | 0.63 | 0.51 | - | 3.05 |
| Burundi | 0.28 | 0.17 | 0.20 | 0.01 | 0.66 |
| Cambodia | 0.92 | 15.97 | 2.99 | - | 19.89 |
| Cameroon | 2.10 | 0.51 | 1.23 | 0.00 | 3.84 |
| Canada | 13.98 | - | - | 28.98 | 42.97 |
| Cape Verde | 0.00 | - | 0.02 | - | 0.02 |
| Central African Republic | 0.09 | 0.02 | 0.13 | - | 0.24 |
| Chad | 0.36 | 0.34 | 0.41 | 0.00 | 1.12 |
| Chile | 0.79 | 0.20 | - | 1.76 | 2.76 |
| China | 272.76 | 300.17 | 107.26 | 178.04 | 858.22 |
| Colombia | 1.59 | 4.66 | 24.03 | 0.01 | 30.29 |
| Comoros | 0.01 | 0.04 | - | - | 0.05 |
| Congo | 0.01 | 0.00 | 0.72 | - | 0.73 |
| Costa Rica | 0.01 | 0.21 | 4.30 | - | 4.51 |
| Côte d'Ivoire | 1.14 | 2.32 | 2.10 | - | 5.56 |

| Country | Maize | Rice | Sugarcane | Wheat | Total |
|---|---|---|---|---|---|
| Croatia | 2.24 | 0.00 | - | 1.28 | 3.53 |
| Cuba | 0.24 | 0.32 | 11.21 | - | 11.76 |
| Cyprus | - | 0.00 | - | 0.03 | 0.03 |
| Czech Republic | 0.99 | 0.00 | - | 6.45 | 7.44 |
| D. P. R. of Korea | 2.30 | 2.60 | - | 0.11 | 5.01 |
| Democratic Republic of Congo | 2.24 | 2.21 | 2.18 | 0.01 | 6.65 |
| Denmark | 0.05 | 0.00 | - | 5.26 | 5.31 |
| Djibouti | 0.00 | - | 0.00 | - | 0.00 |
| Dominica | 0.00 | - | 0.00 | - | 0.01 |
| Dominican Republic | 0.05 | 1.41 | 5.51 | - | 6.97 |
| Ecuador | 1.70 | 2.11 | 11.37 | 0.01 | 15.19 |
| El Salvador | 0.88 | 0.03 | 7.51 | - | 8.42 |
| Eritrea | 0.02 | - | - | 0.03 | 0.05 |
| Estonia | 0.00 | 0.00 | - | 0.96 | 0.96 |
| Eswatini | 0.10 | 0.00 | 5.72 | 0.00 | 5.82 |
| Ethiopia | 10.72 | 0.28 | 1.17 | 6.78 | 18.95 |
| Fiji | 0.00 | 0.01 | 1.42 | - | 1.43 |
| Finland | 0.00 | 0.00 | - | 0.89 | 0.89 |
| France | 15.36 | 0.09 | - | 47.53 | 62.97 |
| Gabon | 0.05 | 0.00 | 0.29 | - | 0.34 |
| Georgia | 0.23 | - | - | 0.18 | 0.41 |
| Germany | 4.46 | 0.00 | - | 27.90 | 32.36 |
| Ghana | 3.50 | 1.72 | 0.15 | - | 5.38 |
| Greece | 1.32 | 0.34 | - | 1.38 | 3.04 |
| Grenada | 0.00 | - | 0.01 | - | 0.01 |
| Guatemala | 1.96 | 0.05 | 27.76 | 0.00 | 29.76 |
| Guinea | 0.80 | 3.47 | 0.32 | - | 4.58 |
| Guinea-Bissau | 0.02 | 0.30 | 0.01 | - | 0.32 |
| Guyana | 0.00 | 0.78 | 1.28 | - | 2.07 |
| Haiti | 0.20 | 0.22 | 1.47 | - | 1.88 |
| Honduras | 0.67 | 0.08 | 4.76 | 0.00 | 5.51 |
| Hong Kong, SAR | - | 0.00 | - | - | 0.00 |
| Hungary | 6.42 | 0.01 | - | 6.88 | 13.32 |
| India | 31.65 | 273.60 | 405.40 | 142.47 | 853.11 |
| Indonesia | 20.01 | 76.18 | 32.20 | - | 128.39 |
| Iraq | 0.37 | 0.59 | 0.00 | 5.50 | 6.47 |
| Ireland | 0.00 | 0.00 | - | 0.82 | 0.82 |
| Islamic Republic of Iran | 0.32 | 2.23 | 8.26 | 13.12 | 23.93 |
| Israel | 0.07 | - | - | 0.20 | 0.26 |

| Country | Maize | Rice | Sugarcane | Wheat | Total |
|---|---|---|---|---|---|
| Italy | 6.08 | 2.04 | - | 9.48 | 17.61 |
| Jamaica | 0.00 | 0.00 | 0.50 | - | 0.50 |
| Japan | 0.00 | 14.74 | 1.31 | 1.43 | 17.47 |
| Jordan | 0.03 | - | - | 0.04 | 0.06 |
| Kazakhstan | 1.13 | 0.71 | - | 15.36 | 17.19 |
| Kenya | 3.30 | 0.26 | 7.78 | 0.32 | 11.67 |
| Kuwait | 0.01 | - | - | 0.00 | 0.01 |
| Kyrgyz Republic | 0.69 | 0.06 | - | 0.47 | 1.23 |
| Lao People's Democratic Republic | 1.05 | 5.42 | 1.89 | - | 8.35 |
| Latvia | 0.00 | 0.00 | - | 3.13 | 3.13 |
| Lebanon | 0.00 | - | 0.00 | 0.13 | 0.13 |
| Lesotho | 0.09 | - | - | 0.01 | 0.10 |
| Liberia | - | 0.36 | 0.28 | - | 0.63 |
| Libya | 0.00 | - | - | 0.17 | 0.17 |
| Lithuania | 0.10 | 0.00 | - | 5.52 | 5.63 |
| Luxembourg | 0.00 | 0.00 | - | 0.10 | 0.10 |
| Madagascar | 0.23 | 6.15 | 3.12 | 0.00 | 9.51 |
| Malawi | 4.58 | 0.21 | 3.16 | 0.00 | 7.95 |
| Malaysia | 0.07 | 3.39 | 0.02 | - | 3.49 |
| Maldives | 0.00 | - | - | - | 0.00 |
| Mali | 3.60 | 3.39 | 0.65 | 0.03 | 7.66 |
| Malta | 0.00 | 0.00 | - | 0.00 | 0.00 |
| Mauritania | 0.01 | 0.60 | - | 0.01 | 0.62 |
| Mauritius | 0.00 | 0.00 | 2.67 | - | 2.67 |
| Mexico | 27.50 | 0.36 | 55.49 | 4.27 | 87.62 |
| Moldova | 2.79 | 0.00 | - | 2.03 | 4.83 |
| Mongolia | - | - | - | 0.74 | 0.74 |
| Montenegro | 0.00 | - | - | 0.00 | 0.01 |
| Morocco | 0.05 | 0.07 | 0.61 | 9.81 | 10.54 |
| Mozambique | 2.10 | 0.26 | 2.99 | 0.02 | 5.37 |
| Myanmar | 2.30 | 34.87 | 11.65 | 0.13 | 48.95 |
| Namibia | 0.09 | - | - | 0.02 | 0.12 |
| Nepal | 3.00 | 7.87 | 3.18 | 2.77 | 16.82 |
| Netherlands | 0.17 | 0.00 | - | 1.23 | 1.40 |
| New Zealand | 0.21 | 0.00 | - | 0.55 | 0.76 |
| Nicaragua | 0.38 | 0.69 | 7.11 | - | 8.17 |
| Niger | 0.01 | 0.03 | 0.45 | 0.01 | 0.49 |
| Nigeria | 12.75 | 11.68 | 1.50 | 0.12 | 26.04 |
| North Macedonia | 0.13 | 0.03 | - | 0.32 | 0.48 |

| Country | Maize | Rice | Sugarcane | Wheat | Total |
|---|---|---|---|---|---|
| Norway | - | - | - | 0.35 | 0.35 |
| Oman | 0.02 | - | 0.00 | 0.00 | 0.02 |
| Pakistan | 10.63 | 19.58 | 88.65 | 35.70 | 154.57 |
| Panama | 0.14 | 0.57 | 2.41 | - | 3.12 |
| Papua New Guinea | 0.01 | 0.00 | 0.35 | - | 0.37 |
| Paraguay | 4.09 | 1.65 | 7.22 | 1.21 | 14.17 |
| Peru | 1.58 | 4.86 | 9.83 | 0.26 | 16.54 |
| Philippines | 8.30 | 27.94 | 26.28 | - | 62.52 |
| Poland | 7.32 | 0.00 | - | 15.46 | 22.78 |
| Portugal | 0.75 | 0.25 | - | 0.09 | 1.09 |
| Qatar | 0.00 | - | - | 0.00 | 0.00 |
| R. B. De Venezuela | 1.54 | 1.10 | 3.21 | 0.00 | 5.86 |
| Republic of Korea | 0.08 | 7.30 | - | 0.04 | 7.41 |
| Republic of Yemen | 0.04 | - | 0.00 | 0.16 | 0.20 |
| Romania | 14.82 | 0.02 | - | 13.56 | 28.41 |
| Russian Federation | 15.24 | 1.51 | 0.00 | 98.87 | 115.62 |
| Rwanda | 0.48 | 0.18 | 0.10 | 0.02 | 0.78 |
| Saint Kitts and Nevis | - | - | 0.00 | - | 0.00 |
| Saint Lucia | 0.00 | - | 0.00 | - | 0.00 |
| Saint Vincent and the Grenadines | 0.00 | 0.00 | - | - | 0.00 |
| Samoa | - | - | 0.00 | - | 0.00 |
| Saudi Arabia | 0.06 | 0.00 | - | 0.80 | 0.86 |
| Senegal | 0.75 | 1.93 | 1.39 | - | 4.08 |
| Serbia | 6.03 | - | - | 4.48 | 10.50 |
| Sierra Leone | 0.02 | 2.77 | 0.08 | - | 2.87 |
| Singapore | - | - | 0.00 | - | 0.00 |
| Slovak Republic | 1.58 | 0.00 | - | 2.60 | 4.18 |
| Slovenia | 0.39 | 0.00 | - | 0.20 | 0.59 |
| Solomon Islands | - | 0.00 | - | - | 0.00 |
| Somalia | 0.08 | 0.00 | 0.21 | 0.00 | 0.29 |
| South Africa | 16.87 | 0.00 | 17.99 | 2.93 | 37.80 |
| South Sudan | 0.18 | 0.04 | 0.00 | 0.00 | 0.21 |
| Spain | 4.60 | 0.86 | - | 11.13 | 16.60 |
| Sri Lanka | 0.47 | 7.21 | 0.83 | - | 8.51 |
| Sudan | 0.02 | 0.04 | 5.33 | 0.78 | 6.17 |
| Suriname | 0.00 | 0.37 | 0.09 | - | 0.46 |
| Sweden | 0.01 | 0.00 | - | 3.94 | 3.95 |
| Switzerland | 0.10 | - | - | 0.52 | 0.62 |
| Syrian Arab Republic | 0.31 | 0.00 | 0.00 | 2.54 | 2.85 |

| Country | Maize | Rice | Sugarcane | Wheat | Total |
|---|---|---|---|---|---|
| Tajikistan | 0.24 | 0.09 | - | 1.11 | 1.44 |
| Tanzania | 7.04 | 3.76 | 3.52 | 0.09 | 14.41 |
| Thailand | 5.30 | 47.01 | 66.28 | 0.00 | 118.60 |
| The Bahamas | 0.00 | - | 0.06 | - | 0.06 |
| The Gambia | 0.02 | 0.06 | - | - | 0.08 |
| Timor-Leste | 0.06 | 0.06 | - | - | 0.12 |
| Togo | 0.93 | 0.21 | - | - | 1.14 |
| Trinidad and Tobago | 0.01 | 0.00 | 0.00 | - | 0.01 |
| Tunisia | - | - | - | 1.55 | 1.55 |
| Turkey | 6.75 | 1.40 | - | 22.95 | 31.10 |
| Turkmenistan | 0.01 | 0.12 | - | 1.78 | 1.90 |
| Uganda | 2.80 | 0.42 | 5.37 | 0.03 | 8.63 |
| Ukraine | 42.11 | 0.07 | 0.00 | 41.84 | 84.02 |
| United Arab Emirates | 0.02 | - | - | 0.00 | 0.02 |
| United Kingdom | - | - | - | 18.18 | 18.18 |
| United States of America | 383.94 | 12.18 | 29.96 | 58.23 | 484.31 |
| Uruguay | 0.77 | 1.83 | 0.46 | 1.22 | 4.28 |
| Uzbekistan | 0.59 | 0.47 | - | 7.78 | 8.84 |
| Vanuatu | 0.00 | - | - | - | 0.00 |
| Vietnam | 4.45 | 61.39 | 10.74 | - | 76.58 |
| Zambia | 3.62 | 0.09 | 5.10 | 0.27 | 9.08 |
| Zimbabwe | 1.47 | 0.00 | 3.45 | 0.44 | 5.36 |
| World | 1210.24 | 1102.21 | 1859.39 | 1002.14 | 5173.98 |

**Table S3.** Total crop residues produced for maize, rice, sugarcane, and wheat, by country, in million tons. Color scale indicates relative production volume, with green shades representing higher production values.

| Country | Cattle | Horses | Sheep | Swine | Total |
|---|---|---|---|---|---|
| Afghanistan | 5.12 | 0.02 | 13.53 | - | 18.67 |
| Albania | 0.34 | 0.08 | 1.48 | 0.16 | 2.05 |
| Algeria | 1.73 | 0.05 | 31.13 | 0.00 | 32.92 |
| Angola | 5.19 | 0.00 | 1.21 | 3.66 | 10.06 |
| Antigua and Barbuda | 0.00 | 0.00 | 0.01 | 0.01 | 0.02 |
| Arab Republic of Egypt | 2.82 | 0.08 | 2.24 | 0.01 | 5.15 |
| Argentina | 53.42 | 2.46 | 13.35 | 5.48 | 74.70 |
| Armenia | 0.58 | 0.01 | 0.63 | 0.01 | 1.42 |
| Australia | 24.43 | 0.22 | 68.05 | 2.58 | 95.27 |
| Austria | 1.87 | - | 0.40 | 2.79 | 5.06 |
| Azerbaijan | 2.52 | 0.06 | 7.31 | 0.01 | 9.90 |
| Bangladesh | 24.55 | - | 2.13 | - | 26.67 |
| Barbados | 0.01 | 0.00 | 0.01 | 0.02 | 0.05 |
| Belarus | 4.24 | 0.03 | 0.08 | 2.55 | 6.90 |
| Belgium | 2.31 | - | - | 6.04 | 8.35 |
| Belize | 0.10 | 0.01 | 0.02 | 0.02 | 0.15 |
| Benin | 2.62 | 0.00 | 1.00 | 0.57 | 4.19 |
| Bhutan | 0.30 | 0.01 | 0.01 | 0.02 | 0.34 |
| Bolivia | 10.39 | 0.51 | 7.63 | 3.26 | 21.79 |
| Bosnia and Herzegovina | 0.34 | 0.00 | 1.03 | 0.56 | 1.93 |
| Botswana | 1.03 | 0.02 | 0.18 | 0.00 | 1.24 |
| Brazil | 224.60 | 5.78 | 20.54 | 42.54 | 293.46 |
| Brunei Darussalam | 0.00 | - | 0.00 | 0.00 | 0.01 |
| Bulgaria | 0.61 | - | 1.20 | 0.69 | 2.51 |
| Burkina Faso | 10.40 | 0.04 | 11.37 | 2.69 | 24.51 |
| Burundi | 0.95 | - | 0.66 | 0.89 | 2.50 |
| Cambodia | 2.70 | 0.03 | - | 2.07 | 4.80 |
| Cameroon | 6.23 | 0.02 | 3.65 | 1.93 | 11.83 |
| Canada | 11.06 | 0.40 | 0.79 | 14.03 | 26.28 |
| Cape Verde | 0.03 | 0.00 | 0.02 | 0.07 | 0.12 |
| Central African Republic | 4.76 | - | 0.43 | 1.06 | 6.25 |
| Chad | 33.29 | 1.38 | 41.77 | 0.11 | 76.55 |
| Chile | 3.04 | 0.30 | 1.72 | 2.88 | 7.94 |
| China | 60.52 | 3.73 | 186.38 | 454.81 | 705.43 |
| Colombia | 29.30 | 1.60 | 1.81 | 5.80 | 38.51 |
| Comoros | 0.05 | - | 0.03 | - | 0.08 |
| Congo | 0.34 | 0.00 | 0.13 | 0.10 | 0.57 |
| Costa Rica | 1.31 | 0.12 | 0.00 | 0.40 | 1.83 |
| Côte d'Ivoire | 1.79 | - | 2.34 | 0.43 | 4.56 |

| Country | Cattle | Horses | Sheep | Swine | Total |
|---|---|---|---|---|---|
| Croatia | 0.43 | - | 0.65 | 0.97 | 2.05 |
| Cuba | 3.66 | 0.91 | 1.38 | 1.94 | 7.88 |
| Cyprus | 0.08 | - | - | 0.36 | 0.45 |
| Czech Republic | 1.36 | 0.03 | 0.18 | 1.49 | 3.07 |
| D. P. R. of Korea | 0.58 | 0.05 | 0.17 | 2.26 | 3.06 |
| Democratic Republic of Congo | 1.49 | 0.00 | 0.92 | 1.00 | 3.41 |
| Denmark | 1.48 | - | - | 13.15 | 14.63 |
| Djibouti | 0.30 | - | 0.47 | - | 0.77 |
| Dominica | 0.01 | - | 0.01 | 0.01 | 0.03 |
| Dominican Republic | 3.06 | 0.36 | 0.26 | 0.55 | 4.23 |
| Ecuador | 4.07 | 0.19 | 0.53 | 1.05 | 5.84 |
| El Salvador | 0.74 | 0.10 | 0.01 | 0.25 | 1.09 |
| Eritrea | 2.13 | - | 2.44 | - | 4.56 |
| Estonia | 0.25 | - | - | 0.31 | 0.56 |
| Eswatini | 0.61 | 0.00 | 0.04 | 0.04 | 0.69 |
| Ethiopia | 65.72 | 2.19 | 38.61 | 0.04 | 106.56 |
| Fiji | 0.06 | 0.05 | 0.03 | 0.15 | 0.29 |
| Finland | 0.83 | - | 0.13 | 1.09 | 2.05 |
| France | 17.33 | 0.29 | 6.99 | 12.94 | 37.56 |
| Gabon | 0.04 | - | 0.23 | 0.22 | 0.49 |
| Georgia | 0.93 | 0.04 | 0.90 | 0.17 | 2.03 |
| Germany | 11.04 | - | 1.51 | 23.76 | 36.31 |
| Ghana | 1.93 | 0.00 | 5.65 | 0.88 | 8.47 |
| Greece | 0.56 | - | 7.25 | 0.73 | 8.55 |
| Grenada | 0.00 | 0.00 | 0.01 | 0.00 | 0.02 |
| Guatemala | 4.10 | 0.13 | 0.60 | 2.98 | 7.81 |
| Guinea | 8.83 | 0.00 | 3.29 | 0.16 | 12.28 |
| Guinea-Bissau | 0.73 | 0.00 | 0.48 | 0.47 | 1.69 |
| Guyana | 0.10 | 0.00 | 0.13 | 0.01 | 0.24 |
| Haiti | 1.52 | 0.50 | 0.26 | 1.04 | 3.32 |
| Honduras | 2.91 | 0.18 | 0.02 | 0.47 | 3.58 |
| Hong Kong, SAR | 0.00 | 0.00 | 0.00 | 0.11 | 0.11 |
| Hungary | 0.91 | 0.04 | 0.89 | 2.73 | 4.56 |
| India | 193.17 | 0.33 | 74.29 | 8.83 | 276.61 |
| Indonesia | 18.05 | 0.40 | 17.90 | 8.01 | 44.37 |
| Iraq | 2.06 | 0.05 | 6.75 | - | 8.87 |
| Ireland | 6.65 | - | 3.99 | 1.71 | 12.35 |
| Islamic Republic of Iran | 5.34 | 0.13 | 45.27 | 0.00 | 50.74 |
| Israel | 0.53 | 0.00 | 0.52 | 0.16 | 1.22 |

| Country | Cattle | Horses | Sheep | Swine | Total |
|---|---|---|---|---|---|
| Italy | 6.28 | - | 6.73 | 8.41 | 21.42 |
| Jamaica | 0.16 | 0.00 | 0.00 | 0.22 | 0.38 |
| Japan | 3.96 | 0.01 | 0.02 | 9.29 | 13.28 |
| Jordan | 0.08 | 0.00 | 3.09 | - | 3.17 |
| Kazakhstan | 8.19 | 3.49 | 18.60 | 0.78 | 31.05 |
| Kenya | 22.85 | 0.00 | 24.80 | 0.67 | 48.33 |
| Kuwait | 0.03 | 0.00 | 0.75 | - | 0.78 |
| Kyrgyz Republic | 1.75 | 0.55 | 5.54 | 0.03 | 7.86 |
| Lao People's Democratic Republic | 2.26 | 0.03 | - | 4.47 | 6.76 |
| Latvia | 0.39 | 0.01 | 0.09 | 0.33 | 0.82 |
| Lebanon | 0.09 | 0.00 | 0.43 | 0.01 | 0.53 |
| Lesotho | 0.35 | 0.04 | 1.26 | 0.03 | 1.69 |
| Liberia | 0.05 | - | 0.29 | 0.31 | 0.65 |
| Libya | 0.20 | 0.05 | 7.38 | - | 7.62 |
| Lithuania | 0.63 | 0.01 | 0.14 | 0.57 | 1.35 |
| Luxembourg | 0.19 | 0.00 | 0.01 | 0.08 | 0.28 |
| Madagascar | 8.84 | 0.00 | 0.85 | 1.25 | 10.94 |
| Malawi | 2.04 | 0.00 | 0.36 | 7.01 | 9.41 |
| Malaysia | 0.70 | 0.00 | 0.12 | 1.86 | 2.69 |
| Maldives | - | - | - | - | 0.00 |
| Mali | 12.85 | 0.61 | 21.15 | 0.09 | 34.69 |
| Malta | 0.01 | - | 0.01 | 0.04 | 0.07 |
| Mauritania | 1.94 | 0.07 | 11.01 | - | 13.02 |
| Mauritius | 0.00 | 0.00 | 0.00 | 0.02 | 0.03 |
| Mexico | 36.00 | 6.40 | 8.77 | 18.93 | 70.10 |
| Moldova | 0.11 | 0.02 | 0.47 | 0.34 | 0.95 |
| Mongolia | 5.02 | 4.32 | 31.09 | 0.03 | 40.46 |
| Montenegro | 0.08 | 0.00 | 0.18 | 0.02 | 0.29 |
| Morocco | 3.18 | 0.19 | 22.73 | 0.01 | 26.10 |
| Mozambique | 2.22 | - | 0.23 | 1.68 | 4.13 |
| Myanmar | 10.30 | 0.10 | 0.44 | 6.78 | 17.62 |
| Namibia | 2.95 | 0.04 | 1.44 | 0.10 | 4.53 |
| Nepal | 7.47 | - | 0.79 | 1.59 | 9.85 |
| Netherlands | 3.71 | 0.10 | 0.73 | 10.87 | 15.40 |
| New Zealand | 10.15 | 0.04 | 25.73 | 0.25 | 36.17 |
| Nicaragua | 5.66 | 0.27 | 0.01 | 0.56 | 6.49 |
| Niger | 17.11 | 0.26 | 14.13 | 0.04 | 31.54 |
| Nigeria | 21.16 | 0.11 | 48.64 | 8.09 | 78.00 |
| North Macedonia | 0.18 | 0.01 | 0.63 | 0.19 | 1.01 |

| Country | Cattle | Horses | Sheep | Swine | Total |
|---|---|---|---|---|---|
| Norway | 0.90 | 0.04 | 2.26 | 0.77 | 3.97 |
| Oman | 0.42 | - | 0.64 | - | 1.06 |
| Pakistan | 51.50 | 0.38 | 31.60 | - | 83.47 |
| Panama | 1.51 | 0.10 | - | 0.40 | 2.02 |
| Papua New Guinea | 0.09 | 0.00 | 0.01 | 2.12 | 2.22 |
| Paraguay | 13.92 | 0.22 | 0.31 | 1.37 | 15.83 |
| Peru | 5.85 | 0.75 | 10.94 | 3.34 | 20.89 |
| Philippines | 2.61 | 0.25 | 0.03 | 9.94 | 12.82 |
| Poland | 6.38 | 0.16 | 0.27 | 10.24 | 17.04 |
| Portugal | 1.64 | - | 2.24 | 2.22 | 6.10 |
| Qatar | 0.05 | 0.01 | 0.92 | - | 0.98 |
| R. B. De Venezuela | 16.22 | 0.53 | 0.63 | 3.04 | 20.41 |
| Republic of Korea | 3.99 | 0.03 | 0.00 | 11.22 | 15.23 |
| Republic of Yemen | 1.66 | 0.00 | 9.26 | - | 10.92 |
| Romania | 1.83 | - | 10.09 | 3.62 | 15.53 |
| Russian Federation | 18.03 | 1.30 | 19.79 | 25.85 | 64.97 |
| Rwanda | 1.68 | - | 0.55 | 1.38 | 3.62 |
| Saint Kitts and Nevis | 0.00 | - | 0.01 | 0.00 | 0.01 |
| Saint Lucia | 0.01 | 0.00 | 0.01 | 0.01 | 0.03 |
| Saint Vincent and the Grenadines | 0.00 | - | 0.01 | 0.01 | 0.02 |
| Samoa | 0.04 | 0.00 | - | 0.13 | 0.18 |
| Saudi Arabia | 0.70 | 0.00 | 9.37 | - | 10.07 |
| Senegal | 3.68 | 0.58 | 7.73 | 0.47 | 12.46 |
| Serbia | 0.86 | 0.01 | 1.70 | 2.87 | 5.44 |
| Sierra Leone | 0.73 | 0.45 | 1.00 | 0.24 | 2.42 |
| Singapore | 0.00 | - | - | 0.00 | 0.00 |
| Slovak Republic | 0.43 | - | 0.29 | 0.45 | 1.18 |
| Slovenia | 0.48 | - | 0.12 | 0.22 | 0.82 |
| Solomon Islands | 0.02 | 0.00 | - | 0.06 | 0.07 |
| Somalia | 4.44 | 0.00 | 11.41 | 0.00 | 15.85 |
| South Africa | 12.23 | 0.33 | 21.46 | 1.34 | 35.37 |
| South Sudan | 13.61 | 0.00 | 13.99 | - | 27.61 |
| Spain | 6.58 | - | 15.08 | 34.45 | 56.11 |
| Sri Lanka | 1.13 | 0.00 | 0.01 | 0.10 | 1.24 |
| Sudan | 32.03 | 0.79 | 41.01 | - | 73.83 |
| Suriname | 0.04 | 0.00 | 0.01 | 0.03 | 0.08 |
| Sweden | 1.39 | - | 0.35 | 1.37 | 3.11 |
| Switzerland | 1.51 | 0.05 | 0.35 | 1.37 | 3.28 |
| Syrian Arab Republic | 0.87 | 0.01 | 16.78 | 0.00 | 17.66 |

| Country | Cattle | Horses | Sheep | Swine | Total |
|---|---|---|---|---|---|
| Tajikistan | 2.36 | 0.08 | 4.05 | 0.00 | 6.50 |
| Tanzania | 30.72 | - | 6.52 | 0.53 | 37.77 |
| Thailand | 4.63 | 0.01 | 0.04 | 7.74 | 12.42 |
| The Bahamas | 0.00 | 0.00 | 0.01 | 0.01 | 0.01 |
| The Gambia | 0.46 | 0.00 | 0.06 | 0.01 | 0.54 |
| Timor-Leste | 0.22 | 0.05 | 0.04 | 0.25 | 0.55 |
| Togo | 0.46 | 0.00 | 1.66 | 1.14 | 3.25 |
| Trinidad and Tobago | 0.04 | 0.00 | 0.01 | 0.03 | 0.08 |
| Tunisia | 0.63 | 0.06 | 6.24 | 0.01 | 6.94 |
| Turkey | 17.85 | 0.08 | 45.18 | 0.00 | 63.11 |
| Turkmenistan | 2.55 | 0.03 | 14.07 | 0.00 | 16.65 |
| Uganda | 14.62 | - | 2.17 | 2.60 | 19.39 |
| Ukraine | 2.87 | 0.20 | 0.62 | 5.88 | 9.57 |
| United Arab Emirates | 0.10 | 0.00 | 2.08 | - | 2.19 |
| United Kingdom | 9.60 | 0.41 | 32.96 | 5.32 | 48.29 |
| United States of America | 93.79 | 10.67 | 5.17 | 74.15 | 183.77 |
| Uruguay | 11.91 | 0.42 | 6.23 | 0.13 | 18.69 |
| Uzbekistan | 13.54 | 0.26 | 19.33 | 0.05 | 33.18 |
| Vanuatu | 0.10 | 0.01 | - | 0.09 | 0.19 |
| Vietnam | 6.37 | 0.05 | - | 23.53 | 29.95 |
| Zambia | 3.82 | - | 0.19 | 1.29 | 5.29 |
| Zimbabwe | 5.34 | 0.03 | 0.37 | 0.27 | 6.01 |
| World | 1529.30 | 60.19 | 1284.85 | 975.41 | 3849.75 |

**Table S4.** Livestock stock for cattle, horses, sheep, and swine, by country, adapted from [44-47], in millions of animals. Color scale indicates relative stock volume, with green shades representing higher stock values.

| Country | CAPEX | OPEX |
|---|---|---|
| Afghanistan | 4,364,419 | 5,105,032 |
| Albania | 7,275,101 | 2,672,397 |
| Algeria | 2,630,594 | 8,479,947 |
| Angola | 2,630,594 | 4,834,907 |
| Antigua and Barbuda | 9,380,470 | 3,506,976 |
| Arab Republic of Egypt | 2,630,594 | 4,054,212 |
| Argentina | 3,552,901 | 2,799,845 |
| Armenia | 7,275,101 | 2,636,888 |
| Australia | 6,441,559 | 2,455,461 |
| Austria | 7,720,433 | 2,503,933 |
| Azerbaijan | 7,275,101 | 2,485,034 |
| Bangladesh | 4,364,419 | 4,967,913 |
| Barbados | 9,380,470 | 3,204,654 |
| Belarus | 7,275,101 | 2,862,566 |
| Belgium | 6,808,630 | 2,412,765 |
| Belize | 9,380,470 | 3,138,572 |
| Benin | 2,630,594 | 3,709,159 |
| Bhutan | 4,364,419 | 4,199,962 |
| Bolivia | 3,552,901 | 2,910,531 |
| Bosnia and Herzegovina | 7,275,101 | 2,701,317 |
| Botswana | 2,208,828 | 3,626,641 |
| Brazil | 3,607,282 | 2,990,982 |
| Brunei Darussalam | 4,364,419 | 4,890,268 |
| Bulgaria | 7,275,101 | 3,681,531 |
| Burkina Faso | 2,630,594 | 3,760,372 |
| Burundi | 2,630,594 | 3,652,990 |
| Cambodia | 4,364,419 | 5,108,336 |
| Cameroon | 2,630,594 | 3,648,034 |
| Canada | 6,540,000 | 2,540,000 |
| Cape Verde | 2,630,594 | 3,801,673 |
| Central African Republic | 2,630,594 | 3,540,652 |
| Chad | 2,630,594 | 3,717,419 |
| Chile | 3,732,842 | 2,910,432 |
| China | 1,942,120 | 1,551,143 |
| Colombia | 3,318,580 | 3,079,587 |
| Comoros | 2,630,594 | 3,831,409 |
| Congo | 2,630,594 | 3,501,003 |
| Costa Rica | 9,380,470 | 2,678,279 |
| Côte d'Ivoire | 2,630,594 | 3,575,344 |

| Country | CAPEX | OPEX |
|---|---|---|
| Croatia | 7,275,101 | 2,875,151 |
| Cuba | 9,380,470 | 3,172,439 |
| Cyprus | 7,275,101 | 2,377,936 |
| Czech Republic | 7,275,101 | 2,633,883 |
| D. P. R. of Korea | 4,364,419 | 5,006,579 |
| Democratic Republic of Congo | 2,630,594 | 3,532,391 |
| Denmark | 7,275,101 | 2,510,042 |
| Djibouti | 2,630,594 | 3,800,021 |
| Dominica | 9,380,470 | 3,373,161 |
| Dominican Republic | 9,380,470 | 3,097,272 |
| Ecuador | 3,552,901 | 2,639,901 |
| El Salvador | 9,380,470 | 2,751,396 |
| Eritrea | 2,630,594 | 3,367,188 |
| Estonia | 7,275,101 | 2,878,322 |
| Eswatini | 2,630,594 | 3,639,774 |
| Ethiopia | 2,630,594 | 3,426,661 |
| Fiji | 6,063,278 | 9,959,471 |
| Finland | 7,275,101 | 2,414,057 |
| France | 6,834,041 | 2,311,842 |
| Gabon | 2,630,594 | 3,682,726 |
| Georgia | 7,275,101 | 3,049,343 |
| Germany | 8,107,789 | 2,654,405 |
| Ghana | 2,630,594 | 3,125,604 |
| Greece | 7,275,101 | 2,434,458 |
| Grenada | 9,380,470 | 3,275,691 |
| Guatemala | 9,380,470 | 3,097,272 |
| Guinea | 2,630,594 | 3,702,551 |
| Guinea-Bissau | 2,630,594 | 3,818,193 |
| Guyana | 3,552,901 | 3,163,292 |
| Haiti | 9,380,470 | 3,113,792 |
| Honduras | 9,380,470 | 3,108,836 |
| Hong Kong, SAR | 9,609,168 | 4,580,598 |
| Hungary | 7,275,101 | 3,749,375 |
| India | 1,505,009 | 4,545,236 |
| Indonesia | 2,086,684 | 4,256,289 |
| Iraq | 4,364,419 | 4,948,089 |
| Ireland | 8,060,598 | 2,377,661 |
| Islamic Republic of Iran | 4,364,419 | 10,856,519 |
| Israel | 4,364,419 | 4,379,919 |

| Country | CAPEX | OPEX |
|---|---|---|
| Italy | 6,333,296 | 2,322,114 |
| Jamaica | 9,380,470 | 3,201,350 |
| Japan | 8,812,034 | 4,661,126 |
| Jordan | 4,364,419 | 4,539,407 |
| Kazakhstan | 4,364,419 | 10,943,535 |
| Kenya | 1,676,053 | 2,607,934 |
| Kuwait | 4,364,419 | 4,307,913 |
| Kyrgyz Republic | 4,364,419 | 5,113,133 |
| Lao People's Democratic Republic | 4,364,419 | 5,025,735 |
| Latvia | 7,275,101 | 2,650,292 |
| Lebanon | 4,364,419 | 5,022,430 |
| Lesotho | 2,630,594 | 3,596,821 |
| Liberia | 2,630,594 | 4,011,481 |
| Libya | 2,630,594 | 3,575,344 |
| Lithuania | 7,275,101 | 2,785,886 |
| Luxembourg | 7,275,101 | 2,615,674 |
| Madagascar | 2,630,594 | 3,553,868 |
| Malawi | 2,630,594 | 3,651,338 |
| Malaysia | 2,641,454 | 4,328,224 |
| Maldives | 4,364,419 | 5,458,567 |
| Mali | 2,630,594 | 3,601,777 |
| Malta | 7,275,101 | 2,713,145 |
| Mauritania | 2,630,594 | 3,657,946 |
| Mauritius | 2,630,594 | 2,614,509 |
| Mexico | 9,380,470 | 2,825,259 |
| Moldova | 7,275,101 | 2,987,924 |
| Mongolia | 4,364,419 | 4,682,135 |
| Montenegro | 7,275,101 | 2,646,098 |
| Morocco | 2,630,594 | 2,406,004 |
| Mozambique | 2,630,594 | 3,525,783 |
| Myanmar | 4,364,419 | 4,941,481 |
| Namibia | 2,630,594 | 3,616,645 |
| Nepal | 4,364,419 | 5,005,910 |
| Netherlands | 7,168,653 | 2,583,143 |
| New Zealand | 5,684,998 | 22,282,643 |
| Nicaragua | 9,380,470 | 3,316,992 |
| Niger | 2,630,594 | 3,719,071 |
| Nigeria | 3,288,044 | 3,579,100 |
| North Macedonia | 7,275,101 | 2,728,623 |

| Country | CAPEX | OPEX |
|---|---|---|
| Norway | 7,275,101 | 2,461,388 |
| Oman | 4,364,419 | 4,265,540 |
| Pakistan | 4,364,419 | 5,252,521 |
| Panama | 9,380,470 | 3,184,829 |
| Papua New Guinea | 6,063,278 | 11,744,828 |
| Paraguay | 3,552,901 | 2,802,218 |
| Peru | 3,552,901 | 2,877,975 |
| Philippines | 2,399,303 | 4,355,754 |
| Poland | 4,668,775 | 2,917,824 |
| Portugal | 7,275,101 | 2,768,906 |
| Qatar | 4,475,737 | 4,204,151 |
| R. B. De Venezuela | 3,552,901 | 4,666,641 |
| Republic of Korea | 4,726,430 | 4,381,747 |
| Republic of Yemen | 4,364,419 | 4,807,666 |
| Romania | 7,275,101 | 4,289,299 |
| Russian Federation | 4,364,419 | 5,077,349 |
| Rwanda | 2,415,104 | 2,778,841 |
| Saint Kitts and Nevis | 9,380,470 | 3,270,735 |
| Saint Lucia | 9,380,470 | 3,295,516 |
| Saint Vincent and the Grenadines | 9,380,470 | 3,336,816 |
| Samoa | 6,063,278 | 11,903,424 |
| Saudi Arabia | 5,536,589 | 4,471,900 |
| Senegal | 2,630,594 | 2,583,493 |
| Serbia | 7,275,101 | 3,314,491 |
| Sierra Leone | 2,630,594 | 3,664,554 |
| Singapore | 6,681,362 | 2,241,256 |
| Slovak Republic | 7,275,101 | 2,879,235 |
| Slovenia | 7,275,101 | 2,905,521 |
| Solomon Islands | 6,063,278 | 12,450,246 |
| Somalia | 2,630,594 | 3,367,188 |
| South Africa | 2,323,924 | 2,492,950 |
| South Sudan | 2,630,594 | 3,367,188 |
| Spain | 4,709,560 | 2,413,172 |
| Sri Lanka | 4,364,419 | 4,908,833 |
| Sudan | 2,630,594 | 3,406,837 |
| Suriname | 3,552,901 | 2,740,372 |
| Sweden | 6,612,603 | 2,902,849 |
| Switzerland | 10,724,473 | 2,736,779 |
| Syrian Arab Republic | 4,364,419 | 4,936,525 |

| Country | CAPEX | OPEX |
|---|---|---|
| Tajikistan | 4,364,419 | 4,229,661 |
| Tanzania | 2,630,594 | 3,575,344 |
| Thailand | 4,364,419 | 2,053,935 |
| The Bahamas | 9,380,470 | 3,236,042 |
| The Gambia | 2,630,594 | 3,700,899 |
| Timor-Leste | 4,364,419 | 5,194,242 |
| Togo | 2,630,594 | 3,657,946 |
| Trinidad and Tobago | 9,380,470 | 4,520,656 |
| Tunisia | 2,630,594 | 3,494,395 |
| Turkey | 4,364,419 | 12,035,698 |
| Turkmenistan | 4,364,419 | 5,006,579 |
| Uganda | 2,247,051 | 3,637,444 |
| Ukraine | 7,275,101 | 6,769,483 |
| United Arab Emirates | 4,001,471 | 4,979,280 |
| United Kingdom | 9,552,368 | 2,540,665 |
| United States of America | 12,220,940 | 2,719,566 |
| Uruguay | 3,552,901 | 2,738,910 |
| Uzbekistan | 4,364,419 | 4,893,572 |
| Vanuatu | 6,063,278 | 11,898,467 |
| Vietnam | 2,320,080 | 4,277,023 |
| Zambia | 2,630,594 | 3,443,182 |
| Zimbabwe | 4,255,153 | 3,609,897 |
| World Average | 5,327,862 | 3,585,823 |

**Table S5.** Capital expenditure (CAPEX) and yearly operating expenditure (OPEX) for an agricultural pellet plant operating at 40,080 t/y, by country